\documentclass[final,3p,times,singlecolumn,sort&compress]{elsarticle}
\usepackage[colorlinks=true,linkcolor=cyan,citecolor=cyan]{hyperref}

\usepackage{newtxmath,newtxtext}
\usepackage{inputenc}

\usepackage[most]{tcolorbox}
\usepackage{multicol}
\usepackage{subfig}  
\journal{Chinese Journal }
\usepackage{natbib}
\usepackage{graphicx}
\usepackage{bm}
\numberwithin{equation}{section}

\newcommand{\ti}[1]{\ensuremath{ \tilde{#1}}}
\newcommand{\nb}{\ensuremath{ \nabla }}

\newcommand{\m}{\ensuremath{{\mu \nu}}}

\newcommand{\p}{\ensuremath{\partial{}}}

\newcommand{\es}{\ensuremath{\omega_{\rm eff}}}

\newcommand{\om}{\ensuremath{\Omega_{M}}}

\begin{document}

\begin{frontmatter}

\title{Particle Production Scenario in an Algebraically Coupled Quintessence Field with a Dark Matter Fluid}

\author{Saddam Hussain} 
\emailauthor{saddamh@zjut.edu.cn}\\
%
\affiliation{organization={Institute for Theoretical Physics and Cosmology, Zhejiang University of Technology},
            city={Hangzhou},
            postcode={310023}, 
            country={China}}
            
            \date{\today}

\begin{abstract}
We investigate the dynamics of an algebraically coupled quintessence field with a dark matter fluid, focusing on particle production through the action principle via a modified interaction Lagrangian. The interaction parameter serves as the source of dark matter particle production and entropy generation. As particle creation occurs due to the interaction between the field and fluid sectors, the system exhibits additional pressure. Our analysis includes studying the system's dynamics by considering an exponential type of interaction corresponding to the field's exponential potential. We assess the system’s background dynamics using the dynamical system stability technique to derive the constraints on the model parameters. Additionally, we determine the best-fit values of the model parameters against two combinations of data sets: (i) Cosmic Chronometer (CC) + Pantheon + SH0ES, and (ii) CC + Pantheon + SH0ES + SDSS BAO + DESI BAO. By employing a comprehensive data analysis technique, we compare the evidence of our models to flat \(\Lambda\)CDM. Based on the Akaike Information Criterion (AIC) and Bayesian Information Criterion (BIC), one of the models emerges as a robust alternative to \(\Lambda\)CDM when considering the joint data sets.
\end{abstract}


\end{frontmatter}

\section{Introduction}

Over the past two decades, observations have shed light on the dynamics of the cosmos at its largest scale, revealing evidence that the expansion of the universe is accelerating \cite{SupernovaCosmologyProject:1998vns,SupernovaSearchTeam:1998fmf,WMAP:2003elm,Sherwin:2011gv,Wright:2007vr,DES:2016qvw,DES:2021esc,Scolnic:2021amr,Planck:2018vyg}. One widely accepted  explanations for this late-time cosmic acceleration is the existence of an exotic component known as the cosmological constant (\(\Lambda\)), which exerts negative pressure. The cosmological constant accounts for the largest portion of the universe’s energy budget, approximately $70\%$. About $25\%$ of the energy budget is dominated by a non-relativistic, pressureless fluid commonly referred to as dark matter, with the remaining portion composed of baryonic matter.\\
Despite the \(\Lambda\)CDM model's success in explaining most observational evidence, it faces several theoretical challenges, such as the cosmological constant problem, fine-tuning issues, and the cosmological coincidence problem \cite{Copeland:2006wr,Weinberg:1988cp,Rugh:2000ji,Padmanabhan:2002ji,Carroll:1991mt,Arkani-Hamed:2000ifx,Velten:2014nra}. Additionally, the model exhibits discrepancies between the measured values of the Hubble constant \(H_0\) and the amplitude of matter density \(S_8\) between high and low-redshift data, reaching up to the \(4.2 \sigma\) level \cite{DiValentino:2021izs,Krishnan:2020obg,Davis:2019wet,Yang:2018euj,DiValentino:2019jae,Nunes:2022bhn,Vagnozzi:2019ezj}. Consequently, numerous alternatives have been proposed to address these issues, either by modifying the gravitational sector \cite{Sotiriou:2008rp,Nojiri:2006ri,BeltranJimenez:2017tkd,Cai:2015emx,Avilez:2013dxa,SolaPeracaula:2019zsl,SolaPeracaula:2020vpg,Joudaki:2020shz} or by modifying the matter sector \cite{Peebles:2002gy,Nishioka:1992sg,Armendariz-Picon:2000nqq,Armendariz-Picon:2000ulo,Chiba:1999ka,Kamenshchik:2001cp,Bento:2002ps}. In many instances, scalar fields serve as viable candidates for dark energy (DE) and are often minimally coupled with pressureless dark matter (DM) fluid. {However, it is important to note that scalar fields, on their own, do not provide a solution to the longstanding cosmological constant problem. While they can be employed to address or mitigate certain cosmological tensions by effectively mimicking the dynamics found in more fundamental frameworks, the fundamental issue of the smallness of the current dark energy density remains largely unexplained within these models.} Beyond their gravitational signatures, these enigmatic forms of matter pose puzzles to the scientific community. Consequently, numerous possible scenarios have been intensely investigated, including (i) non-gravitational interactions between dark matter and dark energy \cite{Ellis:1989as,Amendola:1999qq,Farrar:2003uw,Zimdahl:2003wg,Sadjadi:2006qp,Hussain:2022dhp,Das:2023rat,Amendola:1999er,Nunes:2000ka,Chimento:2003iea,Zhang:2005rg,Boehmer:2008av,Baldi:2008ay,Leon:2009dt,Majerotto:2009np,Valiviita:2009nu,Cataldo:2010kc,Baldi:2010vv,Leon:2010ai,Clemson:2011an,Fadragas:2014mra,Pan:2019jqh,Gariazzo:2021qtg,Amendola:2003wa} and (ii) non-minimal coupling between matter fields and curvature \cite{Sen:2000zk,Nojiri:2005vv,Bertolami:2008ab,Bertolami:2011rb,Tsujikawa:2006ph,Bettoni:2015wla,Chatterjee:2021ijw,Bhattacharya:2022wzu,Koivisto:2015qua,Hussain:2023kwk,Leon:2008de}.

A recent approach has emerged to investigate the non-gravitational interaction between dark matter (DM) and dark energy (DE) through the variational principle \cite{Boehmer:2015kta,Boehmer:2015sha,Tamanini:2015iia,Kase:2019veo,Gomez-Valent:2020mqn,Goh:2022gxo}. In these studies, dark energy is governed by a scalar field, while the action for the dark matter fluid is modeled using the relativistic fluid action proposed by Brown \cite{Brown:1992kc}. This action encompasses the energy density of the fluid \(\rho\), the particle flux number \(J^{\mu}\), and several Lagrange multipliers. Additionally, a non-gravitational interaction term is introduced at the action level, consisting of fluid and field variables, denoted by \(\mathcal{L}_{\rm int} = - \sqrt{-g} f(n,s,\phi)\), where \(f\) is an arbitrary interaction function depending on the matter degrees of freedom (fluid number density \(n\), entropy per particle \(s\)) and field degrees of freedom \(\phi\). This type of interaction is called algebraically coupled interaction. As shown in ref. \cite{Boehmer:2015kta}, the inclusion of such a coupling leads to an additional interaction energy-momentum tensor appear in Einstein's field equations. Consequently, the individual matter components can no longer conserve independently; however, the total energy-momentum tensor of the matter sector remains conserved. In ref.~\cite{Tamanini:2015iia}, a thorough analysis demonstrated the equivalence of studying the non-gravitational interaction between the dark sector through the variational approach and modifying the matter field equations (the so-called conventional approach) by incorporating a non-vanishing interaction term, \( Q \neq 0 \), i.e., \( \nabla_{\mu} T^{\mu}_{\rm DM} = -Q = -\nabla_{\mu} T^{\mu}_{\phi} \), allowing the transfer of energy between dark matter and dark energy. However, it has been argued that the latter approach may lead to inconsistencies in defining physical quantities, such as the velocity of individual matter components. At the level of cosmological perturbations, this approach can lead to instability in the early-time epoch~\cite{Valiviita:2008iv}.

Any models constructed under the interaction function \( f(n, s, \phi) \) are subject to constraints from the fluid Lagrangian, namely the conservation of number density, \( \nabla_{\mu} (n u^{\mu}) = 0 \), where \( u^{\mu} \) is the fluid's four-velocity, and the conservation of entropy per particle, \( \nabla_{\mu} (s n u^{\mu}) = 0 \). Consequently, when evaluating the covariant derivative of the energy-momentum tensor for the dark matter fluid by decomposing it into components normal and parallel to the fluid's comoving velocity \( u^{\mu} \), the normal component, \( u_{\mu} \nabla_{\nu} T^{\mu\nu}_{\rm DM} = 0 \), vanishes, implying that the energy density scales as \( \rho \propto a^{-3} \), where \( a \) is the scale factor. However, the parallel component does not vanish due to the inclusion of \( \phi \) in the interaction function \( f \), indicating that the system is non-minimally coupled and evolves adiabatically.

In the past two decades, significant research has been devoted to understanding the dynamics of the very early universe when the particle number density is non-conserved \cite{Ford:1986sy,Prigogine:1988jax,Zimdahl:1993cu,Zimdahl:1994rx,Zimdahl:1996ka,Nunes:2016aup,Zimdahl:1997qe,Zimdahl:1999tn,Nunes:2016aup,Bhattacharya:2020bwz,Lima:2014hia,Lima:2015mca,SolaPeracaula:2019kfm}. In these scenarios, the background fluid number density is not conserved, and a source term is artificially introduced into the number density conservation equation, resulting in the generation of creation pressure, \(P_c\). This matter production phenomenon plays a significant role in describing the reheating phase of the inflationary process \cite{10.1093/mnras/266.4.872}. It has also been shown that this effect can significantly contribute to explaining late-time cosmic acceleration~\cite{Lima:2012cm,Cardenas:2020grl}.

In this work, we approach the same problem from a new perspective. As previously stated, a recent approach has been put forth to study non-gravitational interactions from the action principle by introducing an interaction term at the Lagrangian level. To study the matter generation effect from the action principle, we modify the interaction term \(f(n, s, \phi) \rightarrow f(n, s, \phi, \varphi)\), incorporating a Lagrange multiplier \(\varphi\). This modification enables the interaction function \( f \) to serve as the source of particle production. The relativistic fluid action typically includes several Lagrange multipliers; however, we incorporate one fluid scalar field, \( \varphi \), which represents the gradient of chemical free energy, thereby modifying the conservation equations for number density \( \left( \nabla_{\mu} (n u^{\mu}) \neq 0 \right) \) and entropy per particle \( ( \dot{s} \neq 0 ) \). Consequently, the system evolves non-adiabatically in a homogeneous, isotropic, and spatially flat universe.

In this interaction framework, we evaluated thermodynamic relations, such as the temperature of the dark matter fluid. Since \(f\) serves as the source of particle creation, it also exerts creation pressure. Hydrodynamically, this creation pressure can be interpreted as viscous pressure, rendering the background fluid an imperfect fluid. We conducted an in-depth analysis of the background dynamics of this non-minimal coupling scenario using the dynamical system stability technique \cite{Rendall:2001it,Bahamonde:2017ize,Dutta:2017wfd,Coley:2003mj,elias2006critical}, considering an exponential type of interaction \(f \propto \rho^{\beta}e^{\delta \kappa \phi + \gamma \varphi}\) corresponding to the exponential potential for the quintessence field. The choice of exponential interaction function is purely phenomenological, ensuring non-zero effects during both non-accelerating and accelerating regimes. We studied the system's dynamics for the models \(f \propto \rho^{2}\) and \(f \propto \rho^{-1}\). The system yields a late-time accelerating stable phase followed by an attractor matter-dominated phase. Furthermore, we constrain the model parameter space by employing a data analysis technique based on the following data sets: Cosmic Chronometer (CC), Pantheon + SH0ES, the latest SDSS Baryonic Oscillation (BAO) data, and Dark Energy Survey Baryonic Oscillation (DESBAO) points.

The paper is organized as follows: In Sec. \ref{sec:action}, we set up the action for the algebraically coupled field-fluid system and obtain the governing background equations. In Sec. \ref{sec:thermodynamic}, we present brief thermodynamic relations corresponding to the fluid component. A detailed picture of the conservation of the energy-momentum tensor is presented in Sec. \ref{sec:conservation}. The stability of the system is discussed in Sec. \ref{sec:background_dynamics}. In Sec. \ref{sec:data_analysis}, the model parameters are constrained using data analysis. In Sec. \ref{sec:conclusion}, we outline a brief conclusion.

\section{Action for the algebraic interaction \label{sec:action}}

The action describing the algebraically (non-minimally) coupled field-fluid scenario with a modified interaction term is given by:  
\begin{multline}
	S = \int_{\Omega}  d^4x \bigg[ \sqrt{-g} \frac{R}{2\kappa^2} - \sqrt{-g} \rho(n,s) + J^{\mu}(\varphi_{,\mu} + s\theta_{,\mu} + \alpha_{A}{\beta}_{,\mu}^{A} ) - \sqrt{-g}\mathcal{L}_{\phi}(\phi, \p_{\mu}\phi) - \alpha_{1} \sqrt{-g} f(n,s, \phi, \varphi) \bigg] ,
	\label{action_general}
\end{multline}
where, 
\begin{equation}\label{}
	J^\mu = \sqrt{-g}nu^\mu ,  \quad |J|=\sqrt{-g_{\mu\nu}J^\mu J^\nu}, \quad n=\frac{|J|}{\sqrt{-g}}, \quad u^\mu u_\mu=-1\,.
\end{equation}
In this action, the first term corresponds to the Einstein-Hilbert action, where \(g\) denotes the determinant of the metric tensor \(g^{\m}\), \(R\) represents the Ricci scalar, and \(\kappa^2 = 8\pi G\). The second and third terms together represent the action for a relativistic fluid, as introduced by Brown \cite{Brown:1992kc}, where the energy density of the dark matter fluid, \(\rho\), depends on the fluid number density \(n\) and the entropy per particle \(s\).  The other term includes the particle flux density \(J^{\mu}\) and Lagrange multipliers \(\varphi\), \(\theta\), \(\alpha^{A}\), \(\beta_{A}\). Note that Greek indices range from 0 to 3, and \(A\) runs from 1 to 3. It's important to distinguish between \(\alpha_{1}\) and \(\alpha^{A}\), as they are distinct quantities unless specified. Interested readers may refer to recent works such as \cite{Bettoni:2015wla,Andersson:2020phh} for deeper insights into the model constructed using this action. The commas followed by Greek indices indicate covariant derivatives. Before proceeding, we clarify that the action intentionally neglects the contributions from other components of the universe, such as baryonic matter and radiation. Although one can include a fluid Lagrangian for baryonic matter and radiation—either minimally or non-minimally coupled to the scalar field—we have chosen to exclude them for two main reasons.
	
	First, introducing non-minimal couplings between radiation or baryonic fluids and the scalar field would significantly complicate the analysis. This would increase the dimensionality of the phase space, making it difficult to perform the analytical calculations of the fixed points, which are central to our discussion of dark matter and the scalar field in the subsequent sections. Moreover, based on the well-studied $\Lambda$CDM model, it is well established that, in the low-redshift regime, the contributions of radiation and baryonic matter are negligible. Thus, for the late-time evolution, these components can be safely omitted without affecting the validity of the results.
	
	Second, the primary focus of this work is on the background evolution of the model and its comparison with late-time cosmological data, specifically cosmic chronometers, Type Ia supernovae, and late-time BAO data. Since we are not using Cosmic Microwave Background (CMB) or Big Bang Nucleosynthesis (BBN) data—both of which would require considering perturbations, a subject beyond the scope of this paper—the parameters related to the densities of radiation and baryonic matter are not constrained in our analysis. Therefore, for the purposes of this study, we have opted to neglect these contributions.  \\
The field action is given by:
\begin{equation}
	\mathcal{L}_{\phi} = \frac{1}{2}\p_{\mu}\phi \p^{\mu}\phi +V(\phi).
\end{equation}
Here, the Lagrangian represents a standard quintessence field with the potential \(V(\phi)\). The remaining term in the action corresponds to the interaction parameter \(f\), which depends on fluid and field parameters, while \(\alpha_{1}\) is a constant parameter. Taking the variation of the action Eq.~\eqref{action_general} with respect to the metric \(g^{\m}\) yields the Einstein field equation:
\begin{equation}
	R_{\m} -\dfrac{1}{2}R g_{\m} = \kappa^2\left(T_{\m}^{M} + T_{\m}^{\phi} + T_{\m}^{\rm int}\right).
\end{equation} 
Here, the stress tensor of the matter components is defined as \(T_{\m} \equiv \frac{-2}{\sqrt{-g}} \frac{\delta S}{\delta g^{\m}}\). The energy-momentum tensor corresponding to the relativistic fluid is:
\begin{equation}
	T^{\m}_{M} = \rho u^{\mu} u^{\nu} + \left(n\frac{\p \rho}{\p n}- \rho \right) (u^{\mu} u^{\nu}+ g^{\m}).
\end{equation}
Similarly, corresponding to the interaction part: 
\begin{equation}
	T^{\m}_{\rm int} = \alpha_{1} f u^{\mu} u^{\nu} + \alpha_{1} \left(n\frac{\p f}{\p n} - f\right) (u^{\mu} u^{\nu}+ g^{\m}).
\end{equation}
The field's stress tensor becomes:
\begin{equation}
	T_{\phi}^{\m} = -g^{\m} \left(\frac{1}{2} \p_{\alpha}\phi \p^{\alpha} \phi + V(\phi)\right) + \p^{\mu}\phi \p^{\nu} \phi .
\end{equation}
To evaluate the energy density and corresponding pressure of these matter components, we compare with the stress tensor for the perfect fluid $T^{\m} = \rho u^{\mu} u^{\nu} + P (g^{\m} + u^{\mu} u^{\nu})$. Hence, the energy density and pressure of the fluid become:
\begin{equation}
	\rho_{M} = \rho, \quad P_{M} =  \left(n\frac{\p \rho}{\p n}- \rho \right).
\end{equation}
The energy density and pressure corresponding to the algebraic interaction are: 
\begin{equation}
	\rho_{\rm int} = \alpha_{1} f, \quad P_{\rm int} = \alpha_{1} \left(n\frac{\p f}{\p n} - f\right).
\end{equation}
The field energy density and pressure become:
\begin{equation}
	\rho_{\phi} = \frac{1}{2}\dot{\phi}^{2} + V(\phi), \quad P_{\phi} = \frac{1}{2}\dot{\phi}^{2} - V(\phi).
\end{equation}
In the flat FLRW metric $ds^2 = -dt^2 + a(t)^2 d\vec{x}^2$, the Friedmann equations become:
\begin{eqnarray}
	3H^2  & = & \kappa^2(\rho_{M} + \rho_{\phi} +\rho_{\rm int}),\\
	2\dot{H} + 3H^2 & = & - \kappa^2 (P_{M} + P_{\phi} + P_{\rm int}).
\end{eqnarray}
Upon taking the variation of the action with respect to $\phi$, the field equation becomes:
\begin{equation}
	\ddot{\phi} + 3 H \dot{\phi}  +\dfrac{dV(\phi)}{d \phi} + \alpha_{1} \frac{\p f}{\p \phi} = 0.
	\label{field_equation_motion}
\end{equation}

\section{Thermodynamic relations \label{sec:thermodynamic}}

In this section, we examine the thermodynamic aspect of the fluid in light of interaction. Upon varying the action with respect to the fluid variables, the corresponding equations of motion become:
\begin{align}
	J^\mu:& \qquad u_{\mu} \left(\frac{\p \rho}{\p n} + \alpha_{1} \frac{\p f}{\p n}\right) +\varphi_{,\mu}+s\theta_{,\mu}+\beta_A\alpha^A_{,\mu} = 0,\label{eq:05}\\
	s:& \qquad -\left(\frac{\partial \rho}{\partial s} + \alpha_{1} \frac{\p f}{\p s}\right)+nu^{\mu}\theta_{,\mu}=0 \,,\label{eq:06}\\
	\varphi:& \qquad \nb_{\mu} (n u^{\mu}) + \alpha_{1} \dfrac{\p f}{\p \varphi}=0 \,,\label{eq:07}\\
	\theta:& \qquad (sJ^\mu)_{,\mu}=0 \,,\label{eq:08}\\
	\beta_A:& \qquad J^\mu\alpha^A_{,\mu}=0 \,,\label{eq:09}\\
	\alpha^A:& \qquad (\beta_AJ^\mu)_{,\mu}=0. \label{eq:10}
\end{align}

To interpret the physical meaning of some of the Lagrange multipliers, consider the case where the interaction parameter is switched off, i.e., \(\alpha_{1} = 0\). In this scenario, Eqs.~\eqref{eq:07} and \eqref{eq:08} yield conserved fluid number density, \(\nabla_{\mu}(n u^{\mu}) = 0\), and conserved entropy density, \(\dot{s} = 0\), in the flat FLRW metric. From Eq.~\eqref{eq:06}, the temperature of the fluid is defined as \(T \equiv \frac{1}{n}\frac{\partial \rho}{\partial s} \big|_n\), the Lagrange multiplier \(\theta\) becomes the potential for the temperature. Similarly, it can be shown using Eqs.~\eqref{eq:05}, \eqref{eq:06}, and \eqref{eq:09} that \(\varphi\) is the potential for chemical free energy \( F \equiv \mu - Ts = \varphi_{,\mu} u^\mu \), where \(\mu = \frac{\partial \rho}{\partial n}\) is the chemical potential.

Due to the modifications in the interaction parameter, which now includes the Lagrange parameter \(\varphi\), the number density of the fluid from Eq.~\eqref{eq:07} is no longer conserved, and the interaction parameter \(f\) acts as a source of fluid particle production. In the flat FLRW background metric, this equation reads as: 
\begin{equation}
	\dot{n} + 3n H  = - \alpha_{1} \frac{\p f}{\p \varphi}.
	\label{number_density}
\end{equation} 
Here, we have included the Lagrange multiplier \(\varphi\) in the interaction function, but one may also add \(\theta\) dependence, resulting in the interaction function \(f(n,s,\phi, \varphi, \theta)\). The inclusion of \(\theta\) only modifies Eq.~\eqref{eq:08}, and the corresponding equation of motion becomes:
\begin{equation}
	\nabla_{\mu}(s n u^{\mu}) = -\alpha_{1} \frac{\partial f}{\partial \theta}.
\end{equation}
Now, using Eq.~\eqref{number_density}, the entropy evolution equation becomes:
\begin{equation}
	\dot{s} = \frac{\alpha_{1}}{n}\left(s \frac{\partial f}{\partial \varphi} - \frac{\partial f}{\partial \theta}\right).
\end{equation}
It is important to note that if the interaction function only depends on the temperature potential \(\theta\), the fluid evolves non-adiabatically while the particle number density remains conserved. However, the inclusion of \(\varphi\) in the interaction function can simultaneously generate matter and entropy production in the background fluid. If the interaction function \(f\) depends on both Lagrange multipliers, an interesting phenomenon may occur. The derivatives inside the parenthesis in the above equation take alternate signs, and it is possible for certain model choices that:
\begin{equation}
	\left(s \frac{\partial f}{\partial \varphi} - \frac{\partial f}{\partial \theta}\right) = 0.
\end{equation}
This scenario indicates the adiabatic matter production mechanism. However, for simplicity, we proceed with the interaction function as defined in Eq.~\eqref{action_general}. The corresponding contribution to entropy becomes:
\begin{equation}
	\dot{s} = \frac{s}{n} \alpha_{1} \frac{\partial f}{\partial \varphi}.
	\label{entropy_density}
\end{equation}
Hence, the dependence of \(\varphi\) in the interaction function may increase or decrease the fluid particle density, resulting in a change in entropy. Consequently, the evolution of the fluid sector becomes non-adiabatic. As the interaction parameter induces changes in number density and entropy, the corresponding change in energy density of the fluid sector can be explored using the first thermodynamic relation:
\begin{equation}
	d(\rho V) = d Q - P dV.
\end{equation}
Here, \(dQ\) denotes the heat received by the fluid sector during time \(dt\), and \(V= a^3 \) signifies the comoving volume. {For a closed system where the number of particles \(N\) is constant, the above relation can be re-expressed as:}
\begin{equation}
	d(\rho/n) = Tds - P d(1/n).
\end{equation} 
{Here, we use \(n = N/V\), \(s = S/N\), and \(T = \frac{1}{n} \left(\frac{\partial \rho}{\partial s}\right)_n\), which is the temperature of the fluid. However, since particle production occurs in the current system, the system can absorb heat and the number of particles \(N\) changes. In this case, the above relation can be extended to an open system using the enthalpy per unit volume \(h = H/V = \rho + P\), as shown in \cite{Prigogine:1988jax}: }
\begin{equation}
	d(\rho V) = T V (d(sn)-s dn) -P dV + \frac{h}{n} d(nV).
	\label{master_eq_therm}
\end{equation}
{This relation can also be derived by considering the Gibbs relation for a system with a varying number of particles, which induces the chemical potential \(\mu\): 
	\begin{equation}
		T dS = d(\rho V) + P dV - \mu dN. 
	\end{equation}
	Simplifying this, we obtain:
	\begin{equation}
		TV(d(sn) - s dn) = d(\rho V) + P dV - (\mu  + T s)d(nV)
	\end{equation}
	By comparing this with Eq.~\eqref{master_eq_therm}, we find that the specific enthalpy is \(\mathbb{k} \equiv \frac{h}{n} = \mu + T s\). Using the relation \(h = \rho + P\), the entropy per particle can be expressed as \(T n s = \rho + P - \mu\), as shown in \cite{Ivanov:2019van}. To determine the energy evolution of the dark matter fluid, we take the time derivative of Eq.~\eqref{master_eq_therm} and obtain:} 
\begin{equation}
	\dot{\rho} + 3 H (\rho+P) = T s \alpha_{1} \frac{\p f}{\p \varphi} - \alpha_{1} \frac{\rho+P}{n} \frac{\p f}{\p \varphi}.
	\label{evo_fluid_energy_thermo}
\end{equation}
From this, we can express the creation pressure or non-minimally induced pressure \(P_{c}\) as:
\begin{equation}
	P_{c} = \frac{1}{3H} \left( \frac{\rho+P}{n}  -  \frac{s}{n} \frac{\p \rho}{\p s}  \right)\alpha_{1} \frac{\p f}{\p \varphi} .
\end{equation} 
Therefore, as the fluid's particle number density changes, the system can exhibit an additional pressure known as creation pressure. The density evolution can be rewritten as:
\begin{equation}
	\dot{\rho} + 3 H (\rho+P + P_{c}) = 0.
	\label{conserv_dens_thermo}
\end{equation}
From Eq.~\eqref{number_density} and Eq.~\eqref{entropy_density}, it is clear that if \(\alpha_{1}\frac{\partial f}{\partial \varphi} < 0\), the interaction contributes to an increment in the number density of the fluid particles. As a consequence, the change in entropy corresponding to the fluid sector decreases. Although the entropy of the matter sector decreases, it's important to note that the entropy of the system as a whole may not decrease. Hence, the second law of thermodynamics for the entire system remains intact.

Using the above equation, the evolution of the temperature \(T\) of the fluid can be determined. We utilize the total derivative of energy density as:
\begin{equation}
	d\rho(n, T) = \left(\frac{\p \rho}{\p n}\right)_{T}dn + \left(\frac{\p \rho}{\p T}\right)_{n} dT,
\end{equation}
After taking the time derivative and using equations Eq.~\eqref{conserv_dens_thermo} and Eq.~\eqref{number_density}, we obtain:
\begin{equation}
	\dot{T} = \frac{1}{\left(\p\rho/\p T\right)_{n}} \left(-3 H (\rho+ P + P_{c})  - \left(\frac{\p \rho}{\p n}\right)_{T} \dot{n}\right).
\end{equation}
To evaluate \(\left(\frac{\partial \rho}{\partial n}\right)_T\), {we apply the thermodynamic relation for an open system:}
\begin{equation}
	ds =  \frac{1}{nT} d\rho - \frac{\rho +P}{n^2 T} dn,
\end{equation}
and using the partial derivative property, we obtain:
\begin{equation}
	\left(\frac{\p s}{\p n}\right)_T = \frac{1}{n T} \left(\left(\frac{\p \rho}{\p n}\right)_{T} -  \frac{\rho +P}{n}\right), \quad \left(\frac{\p s}{\p T}\right)_n = \frac{1}{n T} \left(\frac{\p \rho }{\p T}\right)_n .
\end{equation}
Further using the property \( \frac{\p^{2} s}{\p T \p n} = \frac{\p^{2} s}{\p n \p T}\), we get:
\begin{equation}
	h = \rho+ P  =  n \left(\frac{\p \rho}{\p n}\right)_T +  T \left(\frac{\p P}{\p T}\right)_n\, .
\end{equation}
Here, \(h\) is the enthalpy per unit volume. Inserting this relation into \(\dot{T}\) and using \(\frac{(\p P/\p T)_n}{(\p \rho/\p T)_n} \equiv (\p P/\p \rho)_n\), we finally obtain:
\begin{equation}
	\dot{T}/T = \frac{s \alpha_{1} \p f/\p \varphi}{(\p \rho/\p T)_n} + \frac{\dot{n}}{n} \left(\frac{\p P}{\p \rho}\right)_n .
	\label{temperature_evo}
\end{equation}
Using Eq.~\eqref{entropy_density}, we recover a relation similar to that obtained in ref. \cite{Ivanov:2019van}. The first term on the right-hand side reflects the contribution from the non-conserved entropy of the fluid. To find the temperature evolution, we choose the model corresponding to the pressureless non-relativistic fluid. The energy density of a non-relativistic ideal gas \( (k_{\rm B} = 1)\) is: 
\begin{equation}
	\rho = M n + \frac{3}{2} n T, \label{ideal_gas_law}
\end{equation}
where \(M\) is the mass of gas particles. For this ideal gas model, the pressure, \(P = n(\partial \rho/\partial n) -\rho =0\), vanishes. Using this relation, the temperature relation simplifies to:
\begin{equation}
	\dfrac{\dot{T}}{T} = \dfrac{2 \dot{s}}{3}.
	\label{temp_evo_ideal}
\end{equation}
This gives the variation of temperature with entropy irrespective of any form of interaction parameter \(f\). Hence the entropy becomes:
\begin{equation}
	s =s_0 + \frac{3}{2} \ln (T/T_0).
\end{equation}
Here, the parameters with subscript \(0\) are integration constants representing the present value of the respective parameters. 

\section{Conservation of energy--momentum tensor \label{sec:conservation}}

In the preceding section, we examined the thermodynamic behavior of the fluid sector and used thermodynamic relations to derive the corresponding evolution of fluid density. In this section, we evaluate the covariant derivative of the total stress-energy tensor to study the energy exchange between the fluid and field sectors. To proceed, we first redefine the matter stress tensor as:
\begin{equation}
	T^{\m}_{A} = (\rho+ \alpha_{1} f) u^{\mu} u^{\nu} + \left(n\frac{\p \rho}{\p n} -\rho +\alpha_{1} n \frac{\p f}{\p n}- \alpha_{1}f\right)(u^{\mu} u^{\nu} + g^{\m}).
\end{equation}
With this redefinition, the Einstein tensor becomes $G_{\m} = \kappa^2 (T_{\m}^{A} + T_{\m}^{\phi})$ \footnote{Here \(A\) denotes a label.}. Upon taking the covariant derivative of the stress tensor, we obtain:
\begin{equation}
	\nb_{\mu}T^{\mu 0}_{A} = \dot{\rho} + \alpha_{1}\dot{f} + 3 H (\rho+ \alpha_{1} f + P_{M} + P_{\rm int}) = Q^{0}.
\end{equation}
Utilizing Eq.~\eqref{number_density} and Eq.~\eqref{entropy_density}, we can express:
\begin{equation}
	\nb_{\mu}T^{\mu 0}_{A} = \dot{\rho} + 3 H (\rho + P_{M}) - \alpha_{1}^2 \frac{\p f}{\p \varphi} \frac{\p f}{\p n} + \alpha_{1} \dot{\phi} \frac{\p f}{\p \phi} + \alpha_{1}^2\frac{\p f }{\p s} \left(\frac{s}{n} \frac{\p f}{\p \varphi}\right) + \alpha_{1} \dot{\varphi} \frac{\p f}{\p \varphi} = Q^{0}.
\end{equation} 
To eliminate $\dot{\varphi}$ from the above equation, we contract Eq.~\eqref{eq:05} with $J^{\mu}$ and then use Eq.~\eqref{eq:09}, we get:
\begin{equation}
	\frac{\p \rho}{\p n} - s \dot{\theta} + \alpha_{1} \frac{\p f}{\p n} = \dot{\varphi},
	\label{varphi_def}
\end{equation}
which, when inserted into the above equation, yields:
\begin{equation}
	\nb_{\mu}T^{\mu 0}_{A} = \dot{\rho} + 3 H (\rho + P_{M}) + \alpha_{1} \dot{\phi} \frac{\p f}{\p \phi} + \alpha_{1} \frac{\p \rho }{\p n} \frac{\p f}{\p \varphi} + \alpha_{1}^2\frac{\p f }{\p s} \left(\frac{s}{n} \frac{\p f}{\p \varphi}\right) - \alpha_{1} s \dot{\theta} \frac{\p f }{\p \varphi},
\end{equation}
The time derivative of the temperature gradient $\dot{\theta}$ can be eliminated by using Eq.~\eqref{eq:06}, resulting in:
\begin{equation}
	\nb_{\mu}T^{\mu 0}_{A} = \dot{\rho} + 3 H (\rho + P_{M}) + \alpha_{1} \dot{\phi} \frac{\p f}{\p \phi} + \alpha_{1} \frac{\p \rho }{\p n} \frac{\p f}{\p \varphi} - \alpha_{1} \frac{s}{n} \frac{\p f }{\p \varphi} \frac{\p \rho}{\p s}.
\end{equation}
Using the definition of the pressure of the fluid $P_{M} + \rho= n \frac{\partial \rho}{\partial n}$ and comparing this with Eq.~\eqref{evo_fluid_energy_thermo}, we obtain:
\begin{equation}
	\nb_{\mu}T^{\mu 0}_{A} = \dot{\rho} + 3 H (\rho + P_{M}+ P_{c}) + \alpha_{1} \dot{\phi} \frac{\p f}{\p \phi} = Q^{0},
\end{equation}
where $P_c = \dfrac{1}{3H} \left(\dfrac{\rho+ P_{M}}{n} \alpha_{1} \dfrac{\partial f}{\partial \varphi} - \alpha_{1} \dfrac{s}{n} \dfrac{\partial f }{\partial \varphi} \dfrac{\partial \rho}{\partial s}\right)$. Considering Eq.~\eqref{evo_fluid_energy_thermo}, the covariant derivative of the fluid sector yields:
\begin{equation}
	\nb_{\mu}T^{\mu 0}_{A} = \alpha_{1} \dot{\phi} \frac{\p f}{\p \phi} = Q^{0}.
\end{equation}
The covariant derivative of the field sector can be determined using Eq.~\eqref{field_equation_motion} as:
\begin{equation}
	\nb_{\mu}T^{\mu 0}_{\phi}  = - \alpha_{1} \frac{\p f}{\p \phi} \dot{\phi} = -Q^{0}.
\end{equation}
This exercise demonstrates that the total energy density of the system remains conserved with energy being exchanged between the fluid and field sectors via the interaction term
\begin{equation}
	Q^{0} = \alpha_{1} \frac{\p f}{\p \phi} \dot{\phi}.
\end{equation}

\section{Background dynamics \label{sec:background_dynamics}}

In this section, we perform a stability analysis of the system using the standard linearization technique. We observed that the interaction function alters not only the field's equation of motion but also the fluid's equation of motion. As the non-minimal coupling parameter \(f\) acts as a source of particle creation and entropy generation, the system transitions to a non-adiabatic state. Therefore, evaluating the system’s stability is essential for understanding this complex model. 

The dynamical system approach offers a mathematical framework for analyzing the physical properties of such non-linear models and for constraining the parameter space through the stability of critical points. This approach is particularly useful when the number of parameters is comparatively large, making it important to determine stability at the background level. To elucidate the background evolution of the system, we define a set of dimensionless dynamical variables as:
\begin{equation}
	x = \frac{\kappa \dot{\phi}}{\sqrt{6}H}, \ y = \frac{\kappa \sqrt{V}}{\sqrt{3} H}, \ z  = \frac{\kappa^2 f}{3 H^2}, \ \Omega_{\phi} = \frac{\kappa^2 \rho_{\phi}}{3 H^2}, \ \om = \frac{\kappa^2 \rho}{3 H^2}.
\end{equation}
Expressing the first Friedmann equation in terms of the dynamical variables yields a constraint on the fractional energy density of the fluid: 
\begin{equation}
	\om =  1 - x^2 - y^2 -\alpha_{1} z,
	\label{frd_const}
\end{equation}
and the effective equation of state (EoS) corresponding to the composite system becomes:
\begin{equation}
	\omega_{\rm eff} = \frac{P_{\rm tot}}{\rho_{\rm tot}}=x^2-y^2+\alpha_{1} (\beta -1) z.
\end{equation}
Here, \(\beta\) denotes the model parameter used in the interaction function \(f\) as defined in Eq.~\eqref{int_function}.
Based on the definition of the dynamical variables, the ranges of the variables are as follows:
\begin{equation}
	-\infty <x< +\infty, \quad 0\le  y< \infty, \quad 	-\infty <z< +\infty, \quad 1>\om >0,\quad 1>\Omega_{\phi} > 0.
	\label{range_norm}
\end{equation}
As the model is non-linear and complex, obtaining an analytical solution for stability analysis with any general form of \(f\) is a cumbersome task. Although the phenomenon of particle production is not new, and the quantum theory for these dark components is still unknown, it is crucial to explore all possible forms of \(f\). However, within the framework of dynamical system analysis, it is preferable to choose the model such that the phase space corresponding to the system remains constrained to 3D. As the number of dynamical variables increases, the analysis becomes increasingly difficult. With this in mind, we adopt an exponential interaction form as previously considered in ref. \cite{Boehmer:2015kta}:
\begin{equation}
	f =  M^{4-4\beta} \rho^{\beta} e^{(\delta \kappa \phi + \gamma \varphi)}.
	\label{int_function}
\end{equation}
Here, we also take an exponential form corresponding to the fluid variable \(\varphi\) to simplify the mathematical complexity. 
 We consider an exponential potential for the quintessence field, as it has been extensively studied in the literature. The potential is given by:
\begin{equation}
	V(\phi) = V_0 e^{\lambda \kappa \phi}.
\end{equation}
{One reason to consider this potential is that the field can exhibit a late-time accelerating solution for a certain range of $\lambda$, as shown in refs. \cite{Copeland:2006wr, Copeland:1997et, Haro:2019peq} for a minimally coupled quintessence field in the presence of a pressureless fluid. Another reason to include this potential in our study is that its derivative with respect to $\phi$ is proportional to the potential itself, eliminating the need to introduce additional variables to close the system. Consequently, the potential does not increase the dimension of the phase space.
}
As the interaction function depends on several field and fluid variables, it is important to evaluate \(\dot{z}\) corresponding to the interaction function Eq.~\eqref{int_function}:
\begin{equation}
	\dot{z} = \frac{\kappa^2}{3 H^2} \left(- 3 H \beta f + \delta \kappa f \dot{\phi} + \gamma f (\rho/n - sT)\right) - z \frac{\dot{H}}{H}.
\end{equation}
Here, we see that the derivative yields field variables \(\dot{\phi}\) which can be re-expressed in terms of predefined dynamical variables. However, the derivative also yields certain thermodynamic variables which cannot be expressed in terms of predefined variables. To address this, we need to define some additional variables corresponding to \((n, s, T)\), as these quantities are time-dependent. The derivative has been evaluated using \(T= \frac{1}{n} \frac{\partial \rho}{\partial s}\), Eq.~\eqref{varphi_def}, and \(\dot{\theta} = T + \alpha_{1} \frac{1}{n} \frac{\partial f}{\partial s}\). Hence, the additional dimensionless dynamical variables are:
\begin{equation}
	\chi = \frac{\kappa^2 n}{ 3 H}, \quad \xi = T/H, \quad s= s.
	\label{dyn_variable_chi}
\end{equation}
We choose the variables corresponding to the number density \(n\), temperature \(T\) of the fluid, and entropy \(s\). With these definitions, the thermodynamic variables are constrained within the following ranges:
\begin{equation}
	\chi >0, \quad \xi >0, \quad s\ge 0.
	\label{thermo_const}
\end{equation}
With these new variables, the total number of independent dynamical variables becomes $6$, i.e., \((x, y, z, \chi, \xi, s)\). As a result, the phase space of the system extends to 6-D. Therefore, for the exponential type of interaction, the phase space of the system cannot be constrained to 3-D. In addition to the extended phase space dimension, this 6-D space is unconstrained, as some of the dynamical variables can take any values. However, in the present study, our primary goal is to explore the stability of the model during the matter-dominated and current epochs of the universe. It is possible that the study of the very early phase of the universe may result in pathological solutions corresponding to these unconstrained variables. In such cases, one must adopt the Poincare compactification for those dynamical variables \cite{ELIAS2006305}. 

In terms of the additionally defined variables, the aforementioned dynamical equation for \(z'\) becomes:
\begin{equation}
	z' = \frac{\dot{z}}{H} = - 3 \beta z + \delta z x \sqrt{6} + \gamma z (\om/\chi - s  \ \xi) + \frac{3}{2} z (1 +x^2 -y^2 + (\beta -1) z \alpha_{1}).
\end{equation} 
By expressing \((\dot{\phi}\), \(\dot{n}\), \(\dot{s}\), and \(\dot{T})\) in terms of the dynamical variables \((x, y, z, \chi, \xi, s)\), we can now write the complete set of autonomous equations governing the system.
The autonomous system of equations that describes the dynamics of the entire system are: 
\begin{eqnarray}
	x' & = & -3 x - \frac{3 \lambda y^2}{\sqrt{6}} - \frac{\delta \alpha_{1} 3 z}{\sqrt{6}} + \frac{3}{2} x (1 +x^2 -y^2 + (\beta -1) z \alpha_{1}) ,\label{x_prime}\\
	y' & = & \dfrac{\sqrt{6} y x \lambda}{2} + \frac{3}{2} y (1 +x^2 -y^2 + (\beta -1) z \alpha_{1}), \label{y_prime}\\
	z' &= &  - 3 \beta z + \delta z x \sqrt{6} + \gamma z (\om/\chi - s  \ \xi) + \frac{3}{2} z (1 +x^2 -y^2 + (\beta -1) z \alpha_{1}), \label{z_prime}\\
	\chi' & = & - \alpha_{1} \gamma z - 3 \chi + \frac{3}{2} \chi (1 +x^2 -y^2 + (\beta -1) z \alpha_{1}), \label{chi_prime}\\
	\xi' & =& \frac{2 \xi s \alpha_{1} z \gamma}{ 3 \chi} + \frac{3}{2} \xi (1 +x^2 -y^2 + (\beta -1) z \alpha_{1}), \label{xi_prime}\\
	s' &= & \frac{s \alpha_{1} z \gamma }{\chi}. \label{s_prime}
\end{eqnarray}
Here, we obtained the dynamics corresponding to the temperature parameter using Eq.~\eqref{temperature_evo}, where we evaluated \((\partial \rho / \partial T)_n\) by considering the ideal gas model given in Eq.~\eqref{ideal_gas_law}. Due to the non-adiabatic particle production mechanism, the degrees of freedom to describe the dynamics of the system increase to 6-dimensions. It's important to note that although we have assumed dark matter to behave as an ideal gas, with its energy density given by Eq.~\eqref{ideal_gas_law}, this alone cannot provide a constraint on either \(\chi\) or \(\xi\). An additional variable is needed to capture the dynamics of \(H\) to close the autonomous system of equations. As a result, the dimension of the phase space remains 6-D. Hence, keeping \(\chi\) and \(\xi\) as dynamical variables reduces the complexity of obtaining the critical points as no logarithmic function will appear. The autonomous equations in \((y', z', \xi', s')\) are invariant under the transformations \((y \mapsto -y, z \mapsto -z, \xi \mapsto -\xi , s \mapsto -s)\), presenting an invariant sub-manifold at \((y=z= \xi=s=0)\). This implies that no trajectory originating in \((y,z,\xi, s \geq 0)\) can cross the \((y=z=\xi=s = 0)\) plane.

In order to obtain the physical properties of the model and constrain the model parameters, we determine the critical points by equating the right-hand side of the autonomous equations to zero. Here, we can see that the differential equation corresponding to \(s'\) imposes stringent constraints on \(z\), as the critical points can only be obtained when either \(s = 0\) or \(z = 0\). In addition, we observe that both \(s'\) and \(z'\) diverge for \(\chi \to 0\). This indicates that either \(n \to 0\) or \(H \gg 1\). As \(n \to 0\) may represent the future epoch of the universe, whereas \(H \gg 1\) indicates the past epoch of the universe. Therefore, to regularize the system corresponding to these critical points, we redefine the time variable as suggested in \cite{Bahamonde:2017ize}:
\begin{equation}
	dN \to \chi d\tilde{N}.
\end{equation}
Since the number of e-folds \(N\) is a monotonically increasing function, the newly defined variable \(\tilde{N}\) also becomes a monotonically increasing function, given that the dynamical variable \(\chi > 0\) is strictly assumed to be positive in this context. By employing this redefinition, we ensure the regularity of the system and facilitate the stability analysis of the model.

As Eq.~\eqref{s_prime} yields the critical point when \(s=0\) or \(z=0\), we consider the case where \(\left(z= \frac{\kappa^2 f}{3 H^2} = 0\right)\) and \(s \ne 0\) as one of the coordinates of the critical points. This indicates that either \(f\) vanishes or \(H \gg 1\). If \(f \rightarrow 0\), the corresponding particle production freezes, as indicated by Eq.~\eqref{number_density}. Consequently, \(\dot{s}\) vanishes, and the system evolves adiabatically with conserved number density. This particular case may be realized in the current or late-time epoch of the universe, where the rate of particle production may decrease due to expansion.

On the other hand, if \(s = 0\) serves as one of the coordinates of the critical point and \(z \ne 0\), Eq.~\eqref{number_density} and Eq.~\eqref{entropy_density} imply that particle production takes place adiabatically as \(\dot{s}\) vanishes. The temperature evolution from Eq.~\eqref{temp_evo_ideal}, corresponding to the dark matter fluid (considered as a pressureless ideal gas), freezes. Hence, the production of dark matter particles may not alter the temperature of the universe. This scenario signifies the  matter and late-time state of the universe.

Now we shall determine the critical points corresponding to the autonomous equations, Eq.~\eqref{x_prime}--Eq.~\eqref{s_prime}, for \(s = 0\). The summary of the critical points is tabulated in Tab. [\ref{tab:critic_with_s}] for the redefined time variable. We then tabulate the physical behavior based on the effective equation of state (EoS) $\es$ and fractional energy densities in Tab. [\ref{tab:condition_existence}], where we outline the existence conditions based on the constraints on dynamical variables Eq.~\eqref{thermo_const}. 

We discuss the stability of the critical points by utilizing the standard linearization technique. To do this, we expand the autonomous system of equations, Eq.~\eqref{x_prime}--Eq.~\eqref{s_prime}, using the Taylor series expansion up to the first order at the critical points. Subsequently, a Jacobian matrix \(J_{ij} = \frac{dx_i'}{dx_j}\big|_{(x_{i*})}\) can be constructed. The real parts of the eigenvalues of this matrix indicate the stability of the critical point. If the real parts of the eigenvalues are all positive (negative), the point becomes an unstable (stable) point. For alternate signs of the real parts, the point becomes a saddle point. If any real part vanishes, the linearization technique becomes invalid for determining the stability of that critical point. In such cases, one must apply the center manifold theorem or numerical analysis to determine the asymptotic behavior of the fixed point. 

The model yields 12 critical points summarized in Tab. [\ref{tab:critic_with_s}]. Among these critical points, a few have vanishing \(z\), resulting in the vanishing of some thermodynamic variables. These critical points represent the case where the quintessence field is minimally coupled with the dark matter fluid. The critical points corresponding to this scenario are \(P_{4\mp}\), \(P_{6}\), \(P_{7}\), and \(P_{9}\). The discussion of these critical points can be found extensively in the literature, as substantial investigation has been done in the past on the quintessence field assuming an exponential type potential. Interested readers may refer to \cite{Copeland:2006wr} and references therein. The behavior of the critical points is discussed as follows:
\begin{table}[t]
	\centering

	\begin{tabular}{lccp{2.5cm}p{2.5cm}p{1.4cm}}
		\hline
		\hline
		\multicolumn{6}{c}{$s_{*}=0$}\\
		\hline
		Points & \(x_{*}\) & \(y_{*}\) & \(z_{*}\) & $\chi_{*}$ & $\xi_{*}$   \\
		\hline 
		
		$P_{1\mp}$ & $\frac{-\sqrt{6} + \sqrt{6}\beta \mp  r}{2\delta}$  & 0 & $\frac{3- 3 \beta \pm  p}{\alpha_{1} \delta^2}$ & $\frac{\gamma (-3 + 3\beta \mp  p)}{3 \delta^2}$ & 0\\

		\hline 
		$P_{2\mp}$ & $\frac{-3 + 3 \beta \mp p }{\sqrt{6}\delta }$  & 0 & $\frac{3-3\beta  \pm p}{\alpha_{1} \delta^2 }$  & $ \frac{\gamma ( -3+3\beta  \mp p)}{3 \delta ^2}$ & Any\\

		\hline
		
		$P_{3}$ & 0 & $\frac{\sqrt{\delta }}{\sqrt{(\beta -1) \lambda +\delta }}$ & $-\frac{\lambda }{\alpha_{1} ((\beta -1) \lambda +\delta )}$ & $\frac{\gamma  \lambda }{3 ((\beta -1) \lambda +\delta )}$ & Any  \\

		\hline
		$P_{4\mp}$ & $\mp 1$ & 0  & 0 & 0 & 0  \\
		\hline

		$P_{5}$ & $\frac{\sqrt{\frac{3}{2}} (\beta -1)}{\delta }$ & $ \frac{\sqrt{\frac{\left(-3 \beta ^2+6 \beta +2 \delta ^2-3\right) ((\beta -1) \lambda +\delta )}{\delta ^2}}}{\sqrt{2} \sqrt{(\beta -1) \lambda +\delta }}$ & $\frac{-3 \beta -\delta  \lambda +3}{\alpha_{1} \delta ^2}$ & $\frac{2 \gamma  (3 \beta +\delta  \lambda -3)}{3 \delta  ((\beta -1) \lambda +2 \delta )}$ & 0\\
		\hline
		
		$P_{6}$ & $-\frac{\sqrt{\frac{3}{2}}}{\lambda }$ & $\frac{\sqrt{\frac{3}{2}}}{\lambda }$ & 0 & 0 & 0\\
		\hline
		
		$P_{7}$ & $-\frac{\lambda }{\sqrt{6}}$ & $\sqrt{1-\frac{\lambda ^2}{6}}$ & $0$ & $0$ & $0$\\
		\hline

		$P_{8}$ & Any & Any & $-\frac{x^2+y^2-1}{\alpha_{1}}$ & 0 & Any \\
		\hline
		$P_{9}$ & Any & Any & 0 & 0 & Any\\
		\hline 
		\hline
		
	\end{tabular}
		\caption{The critical points corresponding to autonomous system of equations Eq.~\eqref{x_prime}--Eq.~\eqref{s_prime}. Here $r = \sqrt{6 \beta ^2-12 \beta -4 \delta ^2+6}$ and $p = \sqrt{9 \beta ^2-18 \beta -6 \delta ^2+9}$.}
	\label{tab:critic_with_s}
\end{table}

\begin{table}[t]
	\tiny
	\centering
	
	\begin{tabular}{l |c| c |c |p{2.5cm}|r}
		\hline
		\hline
		\multicolumn{6}{c}{$s_*=0$}\\
		\hline
		Points & $\Omega_{\phi}$&$\Omega_M$ &$\es$ & \((\chi,\xi \ge 0)\) &Stability\\
		\hline
		$P_{1-} $ & $\frac{\left(\sqrt{6 (\beta -1)^2-4 \delta ^2}+\sqrt{6} (-\beta )+\sqrt{6}\right)^2}{4 \delta ^2}$ & $\frac{(\beta -2) \left(\sqrt{9 (\beta -1)^2-6 \delta ^2}-3 \beta +3\right)}{\delta ^2}+2$ & $-1$ &$\gamma<0, \beta \leq 1-\sqrt{\frac{2}{3}} \sqrt{\delta ^2}$\newline $\gamma>0, \beta>\sqrt{\frac{2}{3}} \sqrt{\delta ^2}+1$ &-- \\
		\hline
		
		$P_{1+}$ & $\frac{\left(\sqrt{3 (\beta -1)^2-2 \delta ^2}+\sqrt{3} \beta -\sqrt{3}\right)^2}{2 \delta ^2}$ & $\frac{2 \left(\sqrt{9 (\beta -1)^2-6 \delta ^2}+\delta ^2-3\right)-\beta  \left(\sqrt{9 (\beta -1)^2-6 \delta ^2}+3 \beta -9\right)}{\delta ^2}$ & $-1$ & $\gamma<0, \beta \leq 1-\sqrt{\frac{2}{3}} \sqrt{\delta ^2}$ \newline $\gamma>0, \beta \geq \sqrt{\frac{2}{3}} \sqrt{\delta ^2}+1$ & --\\
		\hline 
		
		$P_{2-}$ & $\frac{\left(\sqrt{9 (\beta -1)^2-6 \delta ^2}-3 \beta +3\right)^2}{6 \delta ^2}$ & $\frac{(\beta -2) \left(\sqrt{9 (\beta -1)^2-6 \delta ^2}-3 \beta +3\right)}{\delta ^2}+2$ & $-1$ & $\gamma<0, \beta \leq 1-\sqrt{\frac{2}{3}} \sqrt{\delta ^2}$, \newline $\gamma>0, \beta \geq \sqrt{\frac{2}{3}} \sqrt{\delta ^2}+1$ & --\\
		\hline
		
		$P_{2+}$ & $\frac{\left(\sqrt{9 (\beta -1)^2-6 \delta ^2}+3 \beta -3\right)^2}{6 \delta ^2}$ & $\frac{2 \left(\sqrt{9 (\beta -1)^2-6 \delta ^2}+\delta ^2-3\right)-\beta  \left(\sqrt{9 (\beta -1)^2-6 \delta ^2}+3 \beta -9\right)}{\delta ^2}$ & $-1$ & $\gamma<0, \beta \leq 1-\sqrt{\frac{2}{3}} \sqrt{\delta ^2}$ \newline $\gamma>0, \beta \geq \sqrt{\frac{2}{3}} \sqrt{\delta ^2}+1$ &  ---\\
		\hline 
		
		$P_{3}$ & $\frac{\delta }{(\beta -1) \lambda +\delta }$ & $\frac{\beta  \lambda }{(\beta -1) \lambda +\delta }$ & $-1$ & $\lambda>0, \gamma>0, \beta >\frac{\lambda -\delta }{\lambda }$ \newline
		$\lambda>0, \gamma<0, \beta <\frac{\lambda -\delta }{\lambda }$ & -- \\
		\hline 
		
		$P_{4\mp}$ & 1& 0 & 1 & Satisfied & Unstable\\
		\hline 
		$P_{5}$ & 1 & $\frac{3 \beta +\delta  \lambda -3}{\delta ^2}$ & $\frac{\lambda -\beta  \lambda }{\delta }-1$ & Fig. [\ref{fig:p5_density}] & Saddle\\
		\hline
		$P_{6}$ & $\frac{3}{\lambda ^2}$ & $1-\frac{3}{\lambda ^2}$ & 0 & Satisfied & Saddle\\
		\hline 
		$P_{7}$ & 1 & 0 & $\frac{1}{3} \left(\lambda ^2-3\right)$ & Satisfied & Saddle \\
		\hline
		$P_{8}$ & $x_{*}^2 + y_{*}^2$ & 0 &$\beta -\beta  \left(x_{*}^2+y_{*}^2\right)+2 x_{*}^2-1$ & $-1+ x_{*}^2+y_{*}^2 <0, \alpha_{1}>0$, \newline
		$ -1+ x_{*}^2+y_{*}^2 >0, \alpha_{1}<0$. & -- \\
		\hline
		$P_{9}$ & $x_{*}^2 + y_{*}^2$ & $1-(x_{*}^2 + y_{*}^2)$ & $x_{*}^2 - y_{*}^2$ & Satisfied & --\\
		\hline
		\hline

	\end{tabular}
	\caption{The conditions for the existence of the fixed points corresponding to Tab. [\ref{tab:critic_with_s}] and their nature. The dashed lines refer to the discussion in the text.}
	\label{tab:condition_existence}
\end{table}

\begin{itemize}
	\item \textbf{Point $P_{1\pm}$:} At these points, the potential parameter \(y\) and the temperature parameter $\xi$ vanish, while the kinetic part of the field \(x\) is non-zero. The temperature parameter vanishes when \(H \gg 1\), indicating the early epoch of the universe. We are avoiding the condition when temperature \(T \to 0\) as it may indicate the future epoch of the universe. This is because the potential of the field vanishes, making the field less suitable for driving late-time cosmic acceleration.
	The fractional energy densities of the field and fluid at these points are summarized in Tab. [\ref{tab:condition_existence}]. The points yield finite energy density for both matter components and exhibits an accelerating solution with an effective equation of state \(\es= -1\). If this point considers to indicate the early-time accelerating scenario, then the field fractional density must dominate over the fluid energy density at that epoch, i.e., \(\Omega_{\phi} \gg \Omega_M\). 
	
	If we impose the constraints \(0.9 < \Omega_{\phi} < 1\) and \(0 < \Omega_M < 0.1\) on both critical points, the corresponding constraints on the model parameters \(\beta\) and \(\delta\) for the points \(P_{1\mp}\) become:
	\[
	\begin{split}
		\text{$P_{1-}$:} \quad & 0.816 \sqrt{\delta ^2}+1.0\leq \beta <0.472 \sqrt{3.0 \delta ^2+5.0}+0.944, (\delta> 24.50, \delta <-24.50),\\
		\text{$P_{1+}$:} \quad & 1.0\, -0.817 \sqrt{\delta ^2}<\beta <1.0\, -0.577 \sqrt{2.0 \delta ^2+3.0}, (\delta >23.23, \delta < -23.24 ).
	\end{split}
	\]
	From this range, it can be concluded that if the model is to describe an early accelerating phenomenon, the corresponding model parameters need to be fine-tuned. When selecting these parameter ranges, it is possible that the other critical points may not exist. Upon evaluating the stability within these ranges, the point consistently exhibits saddle behavior.
	
	\item \textbf{Points $P_{2\mp}$:} This point has similar coordinates to the previous point; however, the temperature parameter can take any value. This point also exhibits dynamics similar to the aforementioned point. After applying similar conditions, i.e., \(0.9 < \Omega_{\phi} < 1\) and \(0 < \Omega_M < 0.1\), the model parameters need to be fine-tuned in \(\beta\) and \(\delta\) as follows:
	\[
	\begin{split}
		\text{$P_{2-}$:} \quad & 0.816 \sqrt{\delta ^2}+1.0 \leq \beta < 0.472 \sqrt{3.0 \delta ^2+5.0} + 0.944, \quad (\delta > 24.50, \delta < -24.49),\\
		P_{2+}: \quad & 1.0 - 0.817 \sqrt{\delta ^2} < \beta < 1.0 - 0.577 \sqrt{2.0 \delta ^2 + 3.0}, \quad (\delta > 23.24, \delta < -23.24).
	\end{split}
	\]
	This point also exhibits saddle-type behavior.
	
	\item \textbf{Point $P_{3}$:} At this point, the kinetic term of the field vanishes, and \(\xi\) can take any value. With a non-zero potential term the effective equation of state becomes \(-1\), representing the late-time state of the universe. At this point, both the field and fluid fractional energy densities are finite. To constrain the model parameters, we argue that if the model drives late-time cosmic acceleration, the field fractional density must dominate over the fluid energy density, i.e., \(0.5 \le \Omega_{\phi} \le 1\) and \(0 < \Omega_M < 0.4\). This constrains \((\lambda, \delta)\) for various choices of \(\beta\) as shown in Fig. [\ref{fig:p3_density_const}]. Note that \((\beta = 0, 1)\) do not follow these constraints and are thus excluded from our analysis.
	
	In addition to the density constraint, it is essential that the point yields real values and corresponding thermodynamic variable \(\chi\) remains positive, as per Eq.~\eqref{thermo_const}. The conditions for this are given in Tab. [\ref{tab:condition_existence}] and are also shown in Fig. [\ref{fig:p3_density_const}].
	
	To further constrain the other model parameters, we evaluate the stability of the critical point by assuming various \(\beta\) values and fixing \(\alpha_{1} = 1, \xi = 1\) as shown in Figs. [\ref{fig:p3_stab_beta2}, \ref{fig:p3_stab_beta3}, \ref{fig:p3_stab_betam1}] for \((\beta = 2, 3, -1)\) respectively. We find that in each case, one of the real parts of the eigenvalues becomes zero, while the rest take negative and positive values. Consequently, stability cannot be established using the linearization technique alone. Hence, to determine the stability of this point, we will employ a numerical technique. We will evolve the system near this fixed point by varying the initial conditions and model parameters and observing the convergence of the family of curves. If the curves do not converge to this point, it will be classified as a saddle or unstable point.
	
	\begin{figure}[h!]
		\centering
		\subfloat[\label{fig:p3_density_const}]{\includegraphics[scale=0.5]{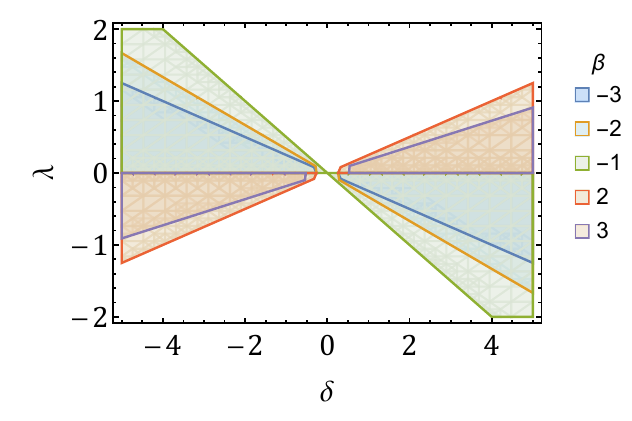}}
		\subfloat[\label{fig:p3_stab_beta2}]{\includegraphics[scale=0.5]{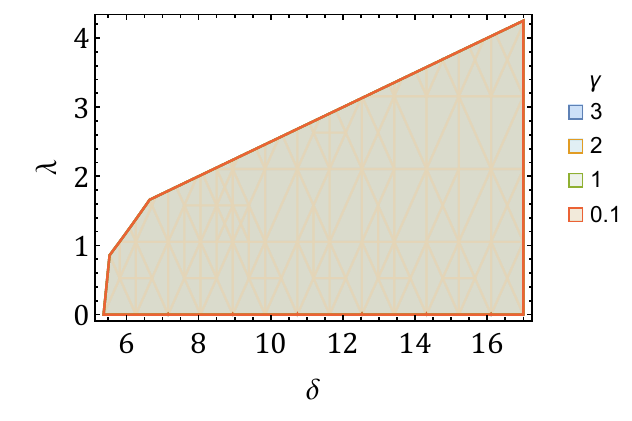}}\\
		\subfloat[\label{fig:p3_stab_beta3}]{\includegraphics[scale=0.5]{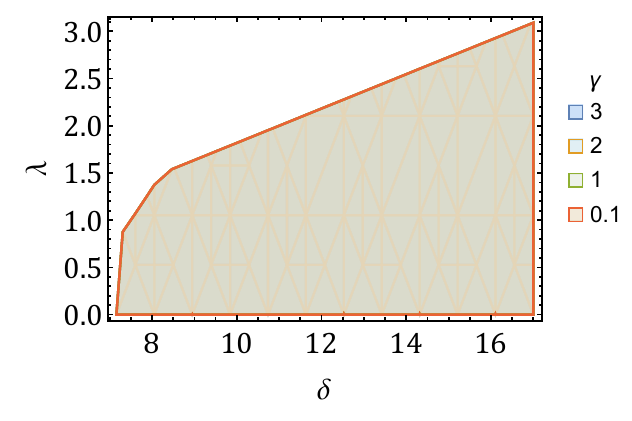}}
		\subfloat[\label{fig:p3_stab_betam1}]{\includegraphics[scale=0.5]{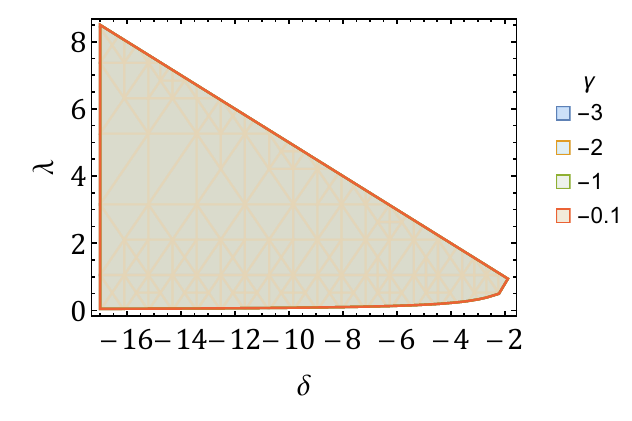}}
		\caption{(a) The region of existence corresponding to the fixed point \(P_{3}\), assuming the constraints \(0.5 \le \Omega_{\phi} \le 1\) and \(0 < \Omega_M < 0.4\) for various choices of \(\beta\), ensuring \(\chi > 0\) and a real value for \(y\) (as it contains a square root term). Here, for \(\beta > 0\), the corresponding \(\gamma > 0\), and for \(\beta < 0\), \(\gamma\) takes negative values. (b) The stability of the critical points evaluated for \(\beta = 2, \alpha_{1} = 1, \xi = 1\), considering all the necessary constraints, where one of the eigenvalues becomes zero. (c) The stability for \(\beta = 3, \alpha_{1} = 1, \xi = 1\), with all the constraints, where one of the eigenvalues becomes zero. (d) The stability for \(\beta = -1, \alpha_{1} = 1,\xi =1\), where one of the eigenvalues becomes zero. Note that due to the overlapping regions, the colors corresponding to different \(\gamma\) values cannot be distinguished. }
		
	\end{figure}
	
	\item \textbf{Points $P_{4\mp}$:} At these points, only the kinetic part of the field is non-zero, and the corresponding field density dominates over the fluid density. Since the effective equation of state is 1, these points exhibit stiff matter characteristics during the very early epoch of the universe. As explained earlier, these points represent the minimal coupling scenario for the quintessence field with an exponential type of potential, as \(z\) vanishes. These points always show unstable behavior.
	
	\item \textbf{Point $P_{5}$:} At this point, all the dynamical variables are non-zero except \(\xi\). Correspondingly, the field density dominates, while the fluid density takes on a finite value depending on the model parameters. The effective equation of state also depends on the model parameters. We can constrain the model parameters \((\delta, \lambda)\) by imposing the conditions \(0 < \Omega_M < 0.4\), \(-1.5 < \omega_{\rm eff} < -1/3\), and \(\chi > 0\). The region of existence is shown in Fig. [\ref{fig:p5_density}] for values of \(\beta\) ranging from negative to positive. Note that \(\beta =0, 1\), do not satisfy the above constraints. For \(\beta > 0\), the point can exhibit both phantom and accelerating solutions, whereas for \(\beta < 0\), it exhibits only an accelerating solution. In both cases, the model parameter \(\gamma\) must be greater than zero. Upon evaluating the stability for any \(\beta\) with the aforementioned constraints, this point consistently exhibits a saddle solution within the confined region. Hence, this point cannot be considered a viable fixed point for representing late-time cosmology.

	\begin{figure}[t]
		\centering
		\includegraphics[scale=0.5]{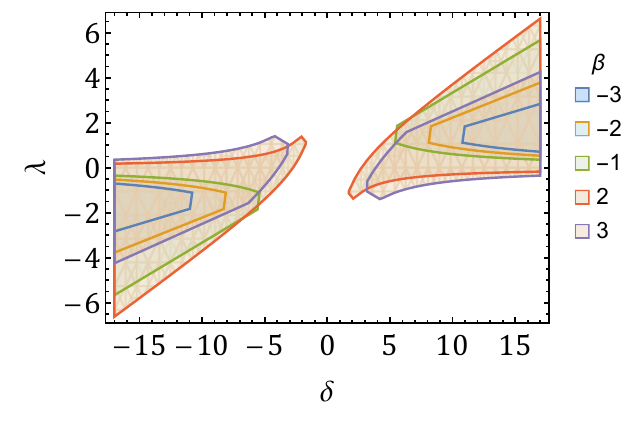}
		
		\caption{ The region of existence for the point $P_{5}$ shown for \(-1.5 < \omega_{\rm eff} < -1/3\) and $0 < \Omega_M < 0.4$, along with $\chi > 0$ and $\gamma > 0$ for negative and positive values of $\beta$. Here, negative $\beta$ shows only accelerating \(-1.0 < \omega_{\rm eff} < -1/3\) solution while $\beta>0$ exhibits both phantom \(-1.5 < \omega_{\rm eff} < -1\) and accelerating  solution for $\gamma>0$. }
			\label{fig:p5_density}
	\end{figure}
	
	\item \textbf{Point $P_{6}$:} This point represents the minimally coupled quintessence case where only the field dynamical variables are non-vanishing. At this point, both the field and fluid fractional densities depend on \(\lambda\). The point follows the constraint relation Eq.~\eqref{range_norm} for \(\lambda^2 > 3\). The effective equation of state is always zero, signifying pressureless fluid type characteristics. The point consistently exhibits saddle-type behavior for any model parameters.

	\item \textbf{Point $P_{7}$:}  Similar to the previous point, the field variables are only dependent on the potential parameter \(\lambda\). At this point, the field density dominates over the fluid density, and the corresponding effective equation of state (EoS) is \(\lambda\) dependent. For \(0 < \lambda^2 < 2\), the model can exhibit an accelerating solution. The point consistently exhibits saddle-type behavior for any choice of model parameters. 
	
	\item \textbf{Point $P_{8}$:} At this point, the coordinates \((x, y, \xi)\) can take any values, constrained by Eq.~\eqref{range_norm}. Consequently, the field energy density can vary between 0 and 1, while the fluid energy density vanishes and remains subdominant. Depending on the dynamical variables and model parameter \(\beta\), this point can exhibit an accelerating solution for \(\beta > 0\). For \(\beta < 0\), the point shows both accelerating and phantom characteristics. This constraint on \((x, y)\) is illustrated in Fig. [\ref{fig:density_p8}] for different values of \(\beta\). The corresponding stability is plotted in Fig. [\ref{fig:p8_stab_betam1}] for \(\beta < 0, \gamma > 0\), as a demonstration corresponding to \((x_* = 0.2, y_* = 0.8)\), ensuring the model satisfies the aforementioned constraint. We found that some of the eigenvalues are zero, positive, and negative. A similar behavior is observed for \(\beta > 0\). Therefore, we need to rely on numerical evolution to determine the stability conclusively.
	
	\begin{figure}[t]
		\centering
		\subfloat[\label{fig:density_p8}]{\includegraphics[scale=0.5]{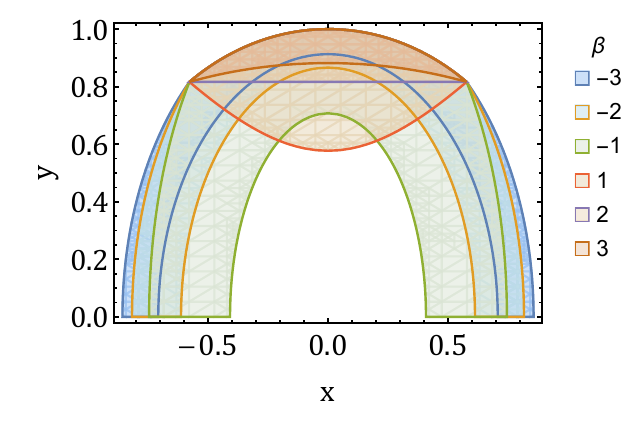}}
		\subfloat[\label{fig:p8_stab_betam1}]{\includegraphics[scale=0.5]{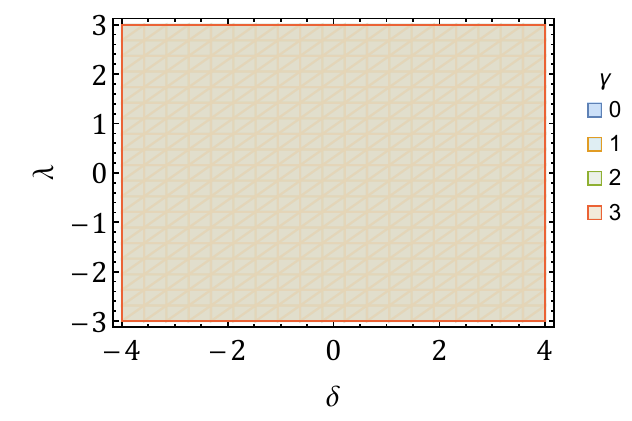}}
		\caption{(a) The region of existence for the point $P_{8}$ with the field energy density \(0 < \Omega_{\phi} < 1\) and \(-1.5 < \omega_{\rm eff} < -1/3\) for different choices of $\beta$. (b) The corresponding stability, where one of the eigenvalues is always zero for \(x_* = 0.2, y_* = 0.8, \beta = -1\), cannot be concluded.}
	\end{figure}
	
	\item \textbf{Point $P_{9}$:} At this point, the dynamical variables corresponding to the field and \(\xi\) can take any value satisfying the constraints in Eqs.~\eqref{range_norm} and \eqref{thermo_const}. The field density can either dominate or sub-dominate depending on the initial conditions and model parameters. The stability of this point cannot be inferred from linear stability analysis alone, necessitating numerical evolution of the system to determine stability conclusively. However, since both the interaction parameter \(z\) and the number density parameter \(\chi\) vanish, this point represents the minimally coupled scenario for the field-fluid model.
	
\end{itemize}

After carefully investigating the nature of the critical points, we found that the critical points where \(s = 0\) represent different epochs of the universe. Some of these critical points, specifically \(P_{1}\) and \(P_{2}\), may exhibit early-time accelerating phenomena. However, the existence conditions of these points indicate the need for fine-tuning of the model parameters \(\beta\) and \(\delta\). Upon evaluating their stability, these points consistently yield saddle behavior.

There are various other critical points where the potential of the field is non-zero and exhibits an accelerating solution. Points \(P_{3}\) and \(P_{8}\), which exert an accelerating solution, may become attractor points. Since the linearization technique cannot be adopted to determine their stability, numerical evolution of the system is necessary to ascertain the behavior of these points. In the subsequent section, we will present the numerical analysis corresponding to \(\beta \gtrless 0\).


\subsection{Model corresponding to $\beta>0$}

\begin{figure}[t]
	\centering
	\includegraphics[scale=0.6]{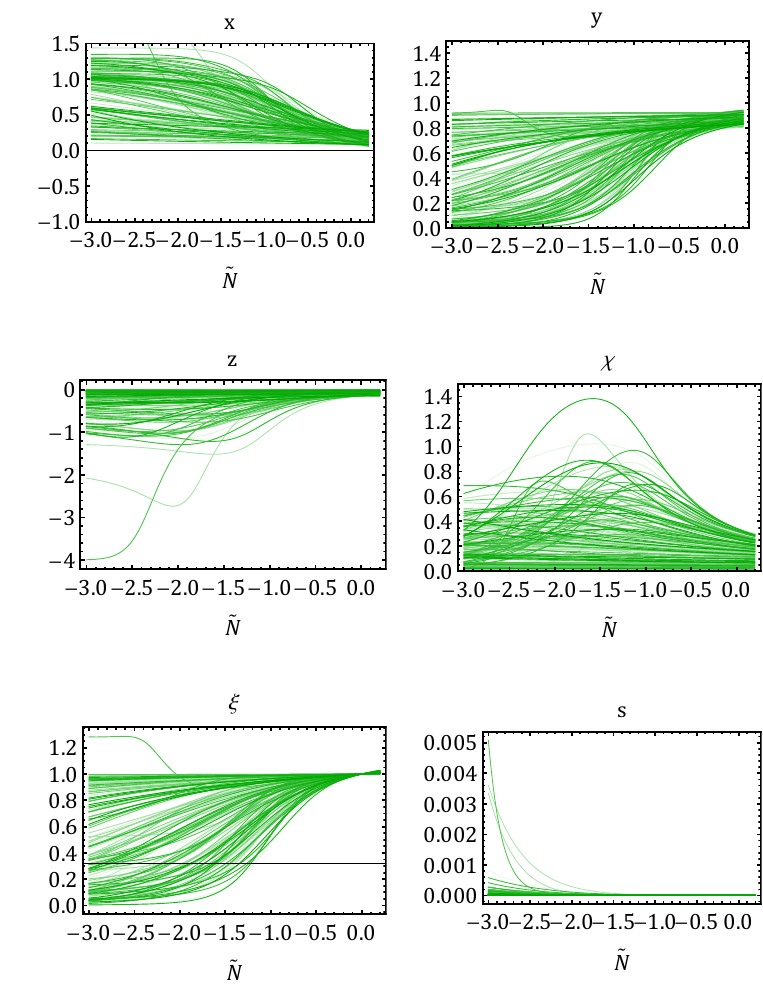}
	\caption{The evolution of dynamical variables corresponding to the various choices of initial conditions and model parameters corresponding to $\beta=2$. }
	\label{fig:evo_phase_dyn_beta2}
\end{figure}

In this section, we study the system's stability and evolution for the interacting model \(f \propto \rho^{2}\), where $\beta =2$. From the previous analysis of the critical points, we find that, the critical points corresponding to the interacting scenario, where \(z\), $\chi$ is finite, the stability can not be inferred through the linearization technique. Therefore, we perform the numerical analysis to identify the stable fixed point. For numerical evolution, we consider only those  parameters for which the accelerating critical points $P_{3,5}$ exist. From the existence relations of $P_{3,5}$ points shown in Figs. [\ref{fig:p3_density_const}, \ref{fig:p5_density}], these points are valid for $\beta\ge 2$. Hence we focus our analysis on $\beta=2$. We set \(\alpha_{1}=1\) for the numerical evolution, as its magnitude  does not affect the system's dynamics. Using the model parameters corresponding to the existence of point \(P_{3} \) as shown in Fig. [\ref{fig:p3_density_const}] we set:
\begin{equation}
	\beta= 2, \alpha_{1} =1, \gamma=2, \delta=0.95, \lambda = 0.3,
	\label{model_para1}
\end{equation}
the coordinates of the fixed point \(P_{3}\) become: \((x = 0, y = 0.87, z = -0.24, \chi = 0.16, \xi = \xi_*, s=0)\) and the corresponding field-fluid densities are: \(\Omega_{\phi} = 0.76, \ \Omega_M = 0.48 \). Within this parameter range, the point \(P_{3}\) does not stabilize. The stability of the point \(P_{8}\) also remains inconclusive, as one of its eigenvalues is zero. Therefore, we numerically evolve the system over the following ranges of model parameters and initial conditions to assess stability. The range is given as:
\begin{equation}
	\begin{split}
		\beta = 2,\,0.08< x_0 < 0.3, \ 0.8 < y_0 < 0.93, \ -0.2 < z_0< -0.003, \ 0.01 < \chi_0 < 0.35, \\ \xi_0 = 1, \  \ 1.5 <\gamma <2, \ 0.1<\lambda<0.9, \ 0.4< \delta< 0.9, \ s_0 = 10^{-5}, \ 66<H_0<76\ . 
	\end{split}
\end{equation} 
Here, the range of model parameters are considered near to Eq.~\eqref{model_para1}. We will then generate the random numbers in this range and evolve the system. Here, we have fixed $(\beta)$ and \((s_0)\). The evolution of dynamical variables are plotted against the redefined time variable \((\ti{N})\) as shown in Fig. [\ref{fig:evo_phase_dyn_beta2}]. Here, we have taken the range of initial conditions at the present epoch \((\ti{N} = 0)\), denoted by the subscript \(0\). From evolution of dynamical variables it can be seen that in the late-time epoch, i.e. \((\ti{N}>0)\), the curves corresponding to the dynamical variables stabilizes to $P_{8}$. From the evolution, one can see that in the past epoch, i.e., \((\ti{N}<0)\), the variable \((z,\chi,s)\) are non-zero, however, in the future epoch, these variables are saturating to 0. Throughout the evolution, we observe that the thermodynamic variables \((\chi, \xi, s)\) are either zero or positive, adhering to the constraint given by Eq.~\eqref{thermo_const} on the system.

\begin{figure}[t]
	\centering
	\includegraphics[scale=0.5]{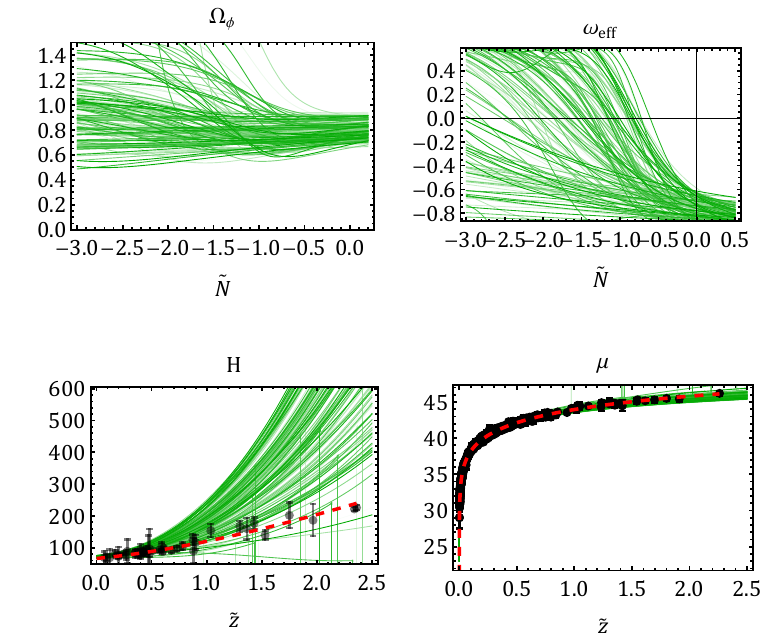}
	\caption{The evolution of cosmological parameters for different initial conditions and model parameters corresponding to $\beta=2$. Here in the \(H(\ti{z})\) evolution, the red dashed line is corresponding to the $\Lambda$CDM model the best fit  \(H_0 = 66.49 \) km/s/Mpc, \(\Omega_{\Lambda} = 0.675\)  obtained in \cite{Brout:2022vxf} for Planck+Pantheon+ (without SH0ES). For the distance modulus \((\mu)\) plot, we have used the SH0ES best fit value for flat $\Lambda$CDM: \(H_0 = 73.6\) km/s/Mpc, $\Omega_{\Lambda} = 0.666$}
	\label{fig:evo_parameter_beta2}
\end{figure}

Correspondingly, we have also plotted the cosmological parameters, for instance, the field density $\Omega_{\phi}$, effective equation of state \(\omega_{\rm eff}\), the Hubble parameter \(H(\ti{z})\) and distance modulus \(\mu(\ti{z})\) in Fig. [\ref{fig:evo_parameter_beta2}]. Here \((\ti{z})\) represents the redshift, \(a = 1/(1+\ti{z})\). To obtain the Hubble evolution, we use the following equation:
\begin{equation}
	\frac{dH}{d \ti{z}} = \frac{3}{2}\frac{1}{(1+ \ti{z})}H(\ti{z} ) (1 + \omega_{\rm eff}).
	\label{hubble_evolution_eq}
\end{equation}
Here, one requires to solve the autonomous equations with respect to the usual time variable i.e., \(N = \ln a\) not \(\ti{N}\). As we are evaluating the dynamics for actual time variable \(N\), however, as long as the system is far from $\chi \to 0$, the system produces similar dynamics as \(\ti{N}\).	The distance modulus is defined as:
\begin{equation}
	\mu  = 5 \log_{10}(d_{L}) + 25 ,
\end{equation}
where the luminosity distance is given by,
\begin{equation}
	d_{L} = 2.99\times 10^{5}(1+\ti{z}) \int_{0}^{\ti{z}}\frac{1}{H(\ti{z})} d\ti{z}.
\end{equation}
For both the Hubble and distance modulus evolution, we have represented the plots against the (43) observational Hubble data and (1701) pantheon+ data (\href{https://github.com/PantheonPlusSH0ES/DataRelease/tree/main/Pantheon%2B_Data/4_DISTANCES_AND_COVAR}{url}) datasets \cite{Cao:2021uda,Perivolaropoulos:2023iqj}. The deviation of the current model has been shown by comparing the curves with the best fit value obtained for $\Lambda$CDM model. 
From the evolution of cosmological parameters shown in Fig. [\ref{fig:evo_parameter_beta2}], we observe that the effective equation of state in the current epoch is in the accelerating regime, with the scalar field density dominating. The Hubble evolution reveals that the curves grow faster compared to $\Lambda$CDM in the higher redshift regime, due to the increase of $\chi$ in the past epoch. The incremental behavior of $\chi$ and \(z\) in the past epoch causes the corresponding field energy density to increase, preventing the field component from vanishing entirely. Although the Hubble evolution plot shows apparent deviations, the distance modulus \((\mu)\) plot does not exhibit significant deviations.


\subsection{Model corresponding to $\beta<0$}

In this section, we present a numerical analysis for the interacting model with $\beta = -1$. For this case, we select the model parameters near to the existence of $P_{3}$ and evolve the system. The range of model parameters and initial conditions are given as: 
\begin{equation}
	\begin{split}
		\beta = -1,\,0.08< x_0 < 0.3, \ 0.7 < y_0 < 0.93, \ 0.02 < z_0< 0.3, \ 0.05 < \chi_0 < 0.7, \ \xi_0 = 2, \ \\ \ 0.5 <\gamma <3,\-0.5<\lambda<0.9, \ 0.4< \delta< 3, \ 0.01<s_0 <0.5,\   \ 66<H_0<76.  
	\end{split}
\end{equation} 
\begin{figure}[h!]
	\centering
	\includegraphics[scale=0.6]{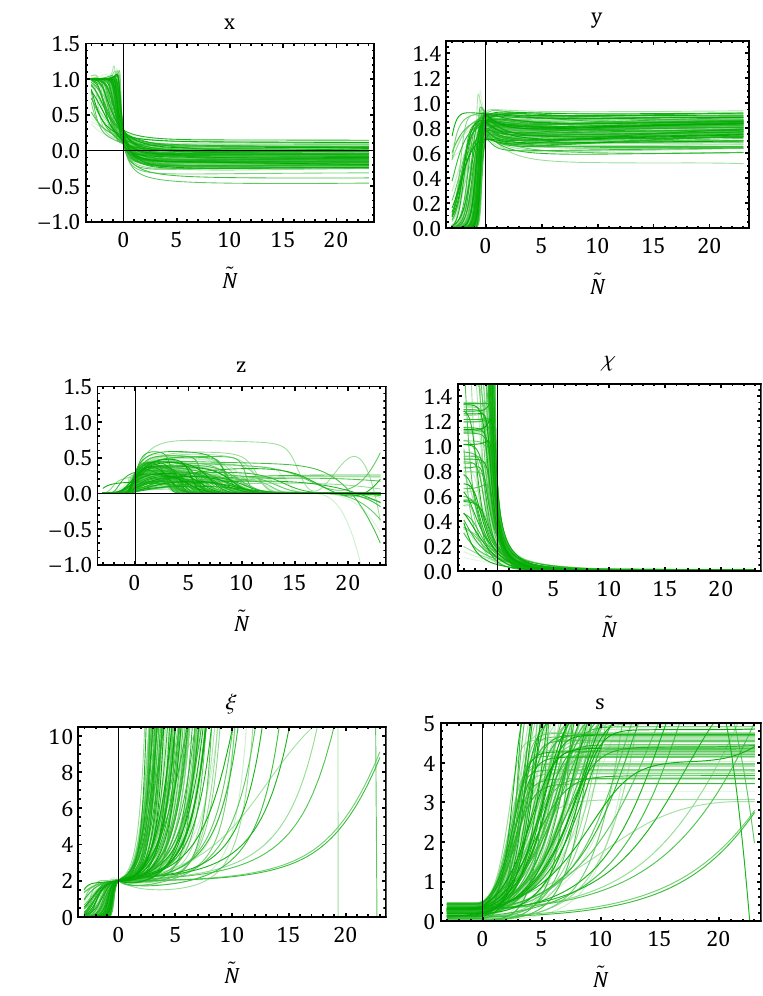}
	\caption{The evolution of dynamical variables for various initial conditions and model parameters corresponding to $\beta=-1$.}
	\label{fig:evo_phase_dyn_betam1}
\end{figure}
\begin{figure}[h!]
	\centering
	\includegraphics[scale=0.6]{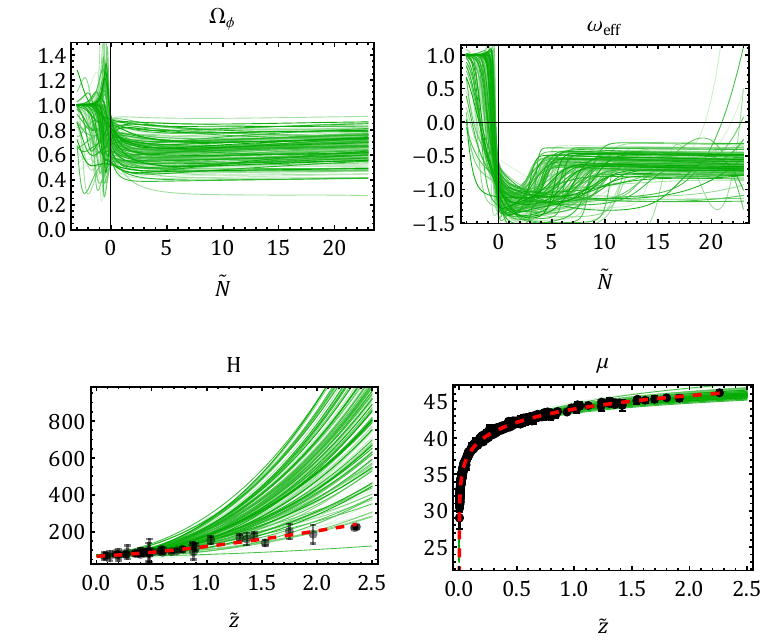}
	\caption{The evolution of cosmological parameters for different initial conditions and model parameters corresponding to $\beta=-1$. In the \(H(\ti{z})\) evolution, the red dashed line corresponds to the $\Lambda$CDM model with the best fit \(H_0 = 66.49\) km/s/Mpc and \(\Omega_{\Lambda} = 0.675\) obtained in \cite{Brout:2022vxf} for Planck+Pantheon+ (without SH0ES). For the distance modulus \(\mu\) plot, we have used the SH0ES best fit value for flat $\Lambda$CDM: \(H_0 = 73.6\) km/s/Mpc and \(\Omega_{\Lambda} = 0.666\).}
	\label{fig:evo_phase_para_betam1}
\end{figure}
The evolution of the dynamical variables is plotted in Fig. [\ref{fig:evo_phase_dyn_betam1}]. From the plot, we observe that in the past epoch, the variable \(z\) vanishes, indicating \(H \gg f\), while \(s\) also vanishes for certain initial conditions and parameter choices. In the current epoch, \(z\) becomes finite but vanishes in the far future, while the entropy saturates at a finite value. The variable \(\chi\) takes a finite value in the past epoch but vanishes in the future. This suggest that in the far future, the particle production ceases as the source term in Eq.~\eqref{number_density} vanishes, and the entropy per particle density becomes constant. Throughout the evolution, \(\xi\) remains positive and eventually saturates in the future. Some abrupt changes are observed for specific combinations of model parameters, which can be discarded. The value of \(x\) is positive in the past but saturates to near zero, both negatively and positively, while \(y\) saturates near \(1\). Hence, the system evolves toward a fixed point where \((z, \chi = 0)\) and \((\xi, s \neq 0)\) approach positive finite values. This behavior closely resembles point \(P_{9}\); however, at this point, both \(s\) and \(z\) are zero. This behavior suggest the existence of an additional critical point, labeled \(P_{10}\), similar to \(P_{9}\), where \(z\) vanishes and \(s\) remains finite and positive. Note that this point is not indexed in Tab. [\ref{tab:critic_with_s}]. The coordinates corresponding to this point are \(P_{10} = (\text{Any}, \text{Any}, 0, 0, \text{Any}, \text{Any})\). The dynamical behavior shows that within a certain range of model parameters, this point becomes stable.

The evolution of cosmological parameter is presented in Fig. [\ref{fig:evo_phase_para_betam1}]. At the current epoch, the effective equation of state lies in the accelerating regime and transitions to the phantom regime \(\es < -1\) in the future epoch for a brief period. During this period, the interaction variable \(z\) becomes non-zero, and the field density remains constrained to \(0 < \Omega_{\phi} < 1\). As the system evolves into the far future, it re-enters an accelerating phase with \(-1 < \es < -0.5\), and \(z\) once again vanishes. In the past, the field energy density becomes greater than \(1\) for some initial conditions and model parameters, indicating a non-physical scenario, since \(z\) remains close to zero in the past epoch, leading to \(\Omega_m < 0\). Therefore, such benchmark points are not viable.

We discuss the data analysis technique used to obtain the best-fit values of the model parameters in the next section.

{
	\section{Data sets \label{sec:data_analysis} }
	In this section, we discuss the observational data used to constrain the parameter space of the proposed models. The datasets employed in this analysis are outlined as follows:
	\begin{itemize}
		\item \textbf{Cosmic Chronometer (CC) data:} This dataset consists of 31 measurements of the Hubble parameter, obtained using the differential age method--a model-independent approach \cite{Yu:2017iju, Thakur:2023ofa, Cao:2022ugh}. The data spans a redshift range of \(\tilde{z} \in [0.07, 1.965]\). The chi-squared statistic (\(\chi^2\)) for this dataset is calculated using the following expression:
		\begin{equation}
			\chi^2_{\rm CC} = \sum_{i=1}^{31} \left(\frac{H_{i \ \rm obs} - H_{i \  \rm Model}}{\sigma_{i}}\right)^{2},
		\end{equation}
		where \(H_{i \ \rm obs}\) represents the observed Hubble parameter at redshift \(\tilde{z}_i\), \(\sigma_i\) is the measurement error, and \(H_{i \ \rm model}\) is the corresponding model prediction.
		
		\item \textbf{SN data:} This dataset includes the latest SNe Ia distance moduli \((\mu)\) measurements from the Pantheon+ and SH0ES sample, which consists of 1701 light curves of 1550 supernovae observed over the redshift range \(z \in [0, 2.3]\) \cite{brout2022pantheon+, scolnic2022pantheon+,Brout:2022vxf}. The distance modulus corresponding to the model is computed using:
		\begin{equation}
			\mu_{\rm Model} = 5 \log_{10}(d_{L}) + 25,
		\end{equation}
		where the luminosity distance \(d_{L}\) is given by,
		\begin{equation}
			d_{L} =  2.99\times 10^{5}(1+\ti{z}) \int_{0}^{\ti{z}}\frac{1}{H(\bar{z})} d\bar{z}.
		\end{equation}
		To constraint the parameter space, we follow the standard $\chi^2$ analysis technique\footnote{Note that here, $\chi$ is not a dynamical variable defined in Eq.~\eqref{dyn_variable_chi}.} which is defined as: 
		\begin{equation}
			\chi^2_{\rm SN} = -2 \ln(\mathcal{L}) = \Delta \bm{D}^{\bm{T}} C_{\rm stat+sys}^{-1} \Delta \bm{D},
		\end{equation}
		where $\Delta \bm{D}$ is a column matrix of residual of the distance modulus:
		\begin{equation}
			\Delta \bm{D} = 	\mu_{\rm data} - \mu_{\rm Model}.
		\end{equation}
		Here, \(C_{\rm stat+sys} = C_{\rm stat} + C_{\rm sys}\) denotes the combined statistical and systematic $1701 \times 1701$ covariance matrix.
		
		\item \textbf{BAO data:} We utilize Baryon Acoustic Oscillation (BAO) observational data to assess the model's viability at the background level, which serves as a critical test of the model. The dataset consists of 8 data points from the Sloan Digital Sky Survey (SDSS) \cite{eBOSS:2020yzd, dawson2016sdss} and 7 data points from the Dark Energy Spectroscopic Instrument (DESI) \cite{DESI:2024mwx, levi2019dark, moon2023first}. These data points are presented in Tab. [\ref{tab:bao_measurements}]. For this analysis, we refer to the SDSS data as BAO and the DESI observations as DESBAO. 
		
		The chi-squared statistic (\(\chi^2\)) for these datasets is defined as:
		\begin{equation}
			\chi^2_{\rm BAO} =  (\text{Data}_i - \text{Model}_i)^T C_{ij}^{-1} (\text{Data}_j - \text{Model}_j) \ ,
		\end{equation}
		where $C_{ij}$ denotes the covariance matrix. 
		For these datasets, the relevant observables include the comoving angular diameter distance \(D_{M}(\tilde{z})\), defined as:
		\begin{equation}
			D_{M}(\tilde{z}) = \int_{0}^{\tilde{z}} c \frac{dz^\prime}{H(z^\prime)},
		\end{equation}
		where \(c\) is the speed of light and \(H(z^\prime)\) is the Hubble parameter at redshift \(z^\prime\). Another important observable is the Hubble distance, \(D_H(\tilde{z}) = c / H(\tilde{z})\). Additionally, we consider the spherically-averaged distance, also known as the dilation scale, \(D_{V}(\tilde{z})\), which is defined as:
		\begin{equation}
			D_{V}(\tilde{z}) \equiv \left({\tilde{z} \ D_M^2(\tilde{z})  D_H}\right)^{1/3}\ .
		\end{equation} 
		The parameter \(r_d\) denotes the comoving sound horizon--the distance traveled by sound waves from the end of inflation until baryon-photons decoupling. This is also referred to as the BAO characteristic scale and is defined as:
		\begin{equation}
			r_d  = \int_{\tilde{z}_d = 1060}^{\infty} \frac{c_s(z^\prime)}{H(z^\prime)} dz^\prime,
		\end{equation}
		where \(c_s\) denotes the sound speed and \(\tilde{z}_d\) is the redshift of the drag epoch. The sound speed \(c_s\) depends on the baryon and photon densities, and is calculated as:
		\begin{equation}
			c_s(\tilde{z}) = \frac{c}{\sqrt{3(1 + \frac{3 \rho_{B}}{4 \rho_{\gamma}})}}\, .
		\end{equation}
		Typically, the BAO characteristic scale is derived from CMB observations. However, in this study, we treat \(r_d\) as a free parameter and constrain it using the current combination of datasets. To calculate \(\chi^2_{\rm BAO}\), we use the following expression:
		\begin{equation}
			\chi^2_{\rm BAO} = \chi^{2}_{D_M/r_d} + \chi^{2}_{D_H/r_d} + \chi^{2}_{D_V/r_d},
		\end{equation}
		as outlined in \cite{eBOSS:2020yzd,Hussain:2024qrd}.
		
		\begin{table}[t]
			\tiny
			
			\centering
			
			\begin{tabular}{l|c|c|c|c|c|c|c|c}
				\hline
				\multicolumn{9}{c}{BAO measurement} \\
				\hline
				\hline
				
				$\tilde{z}_{\rm eff}$ & 0.15 & 0.38 & 0.51 & 0.70& 0.85 & 1.48 & 2.33& 2.33 \\
				
				\hline
				$D_V(\tilde{z})/r_d$ & $4.47 \pm 0.17$ & & & &$18.33^{+0.57}_{-0.62}$  & &&\\
				\hline
				
				$D_M(\tilde{z})/r_d$ & & $10.23 \pm 0.17$ & $13.36 \pm 0.21$ & $17.86 \pm 0.33$ & & $30.69 \pm 0.80$ & $37.6 \pm 1.9$ & $37.3 \pm 1.7$ \\
				\hline
				
				$D_H(\tilde{z})/r_d$ & & $25.00 \pm 0.76$ & $22.33 \pm 0.58$ & $19.33 \pm 0.53$ & & $13.26 \pm 0.55$ & $8.93 \pm 0.28$ & $9.08 \pm 0.34 $\\
				\hline
				\hline
				\multicolumn{9}{c}{DESBAO measurements}\\
				\hline
				\hline
				$\tilde{z}_{\rm eff}$ & 0.295 & 0.510 & 0.706 & 0.930 & 1.317 & 1.491 & 2.330 & \\
				\hline
				$D_V(\tilde{z})/r_d$ & $7.93 \pm 0.15$ &  & && &$26.07 \pm 0.67$ && \\
				\hline
				
				$D_M(\tilde{z})/r_d$ & & $13.62 \pm 0.25$ & $16.85 \pm 0.32$ & $21.71 \pm 0.28$ & $27.79 \pm 0.69$ & & $39.71 \pm 0.94$ &  \\
				\hline
				
				$D_H(\tilde{z})/r_d$ & & $20.98 \pm 0.61$ & $20.08 \pm 0.60$ & $17.88 \pm 0.35$ & $13.82 \pm 0.42$ &  & $8.52 \pm 0.17$ & \\
				\hline
				\hline
				
			\end{tabular}
			\caption{BAO observation datasets.}
			\label{tab:bao_measurements}
		\end{table}
		
	\end{itemize}
	
	To constrain the model parameters, we perform the following joint chi-squared analysis: \\
	(i) \(\chi_{\rm tot}^{2} = \chi_{\rm CC}^{2} + \chi_{\rm SN}^{2}\), and 
	(ii) \(\chi_{\rm tot}^{2} = \chi_{\rm CC}^{2} + \chi_{\rm SN}^{2} + \chi^{2}_{\rm BAO} + \chi^{2}_{\rm DESBAO}\).
	
	\subsection{Observational Results}
This section presents the parameter constraints obtained from the joint chi-squared analysis. The primary focus of this study is the background evolution of the model, treating \(r_d\) as a free parameter. In the absence of CMB data, simultaneously constraining \(H_0\) and \(r_d\) becomes a challenging task due to the degeneracy between these parameters. 
	
	To mitigate this issue, we incorporate additional datasets--including the CC, SN, and BAO samples--and perform a combined chi-squared analysis. Uniform priors are applied to all parameters as summarized in Tab. [\ref{tab:data_analysis_table1}]. Parameters that remain unconstrained by the data are listed without square brackets in the ``prior" row.

	We use publicly available Markov Chain Monte Carlo (MCMC) Python package emcee, based on Goodman and Weare algorithm to perform Bayesian analysis and estimate the posterior distribution and model evidences \cite{Foreman_Mackey_2013}. The posterior samples are further analyzed using the publicly available GetDist Python package, where 1$\sigma$ and 2$\sigma$ contours are generated \cite{lewis2019getdistpythonpackageanalysing}.  The best-fit parameter values are shown in the triangular plots Figs. [\ref{fig:traingleplot_model1}, \ref{fig:traingleplot_model2}], and summarized in Tabs. [\ref{tab:data_analysis_table1}, \ref{tab:Lcdm_model_stat}], at the 68\% confidence level.
	
	\begin{table}[h!]
	
		\tiny
		\begin{tabular}{l|c|c|c|c|c|c|c|c|c|c|c}
			\hline
			Model & $H_0$ & $x_0$ & $y_0$ & $z_0$ & $\chi_0$ & $\xi_0$ & $s_0$ & $\lambda$ & $\gamma$ & $\delta$ & $r_d$\\
			\hline
			\multicolumn{12}{c}{ {\textbf{Model I: $\beta = 2$}}}\\
			\hline 
			\hline
			Priors: & $[30,100]$ & $[-0.1,0.35]$ & $[0.5,1]$ & $[-0.8,0.9] $ & $[0.0,0.3]$ & $1$ & $0.01$ & $[-1.5, 1.2]$ & $1$& $1.8$ & $[100, 350]$  \\
			\hline
			CC+SN & $72.65 \pm 0.30$ & $0.272^{+0.075}_{-0.013}$ & $0.803^{+0.028}_{-0.019}$ & $-0.032^{+0.20}_{-0.064}$ & $0.023^{+0.016}_{-0.023}$ & $1$ & $0.01$ & $-1.10 ^{+0.10}_{-0.33}$ & $1$  & $1.8$ & $--$\\
			
			\hline 
			CC+BAO+DESBAO+SN & $ 72.61 ^{+0.25}_{-0.32}$ & $0.275^{+0.073}_{-0.059}$ & $0.795 ^{+0.037}_{-0.021}$ & $-0.04^{+0.18}_{-0.11} $ &$0.027^{+0.011}_{-0.026}  $& 1 & $0.01$ & $ -1.05 \pm 0.24$ & $1$ & $1.8$ & $135.63 \pm 0.98$ \\
			\hline
			\multicolumn{12}{c}{\textbf{Model II: $\beta =-1$}}\\
			\hline 
			\hline
			Priors: & $[30,100]$ & $[-0.1,0.35]$ & $[0.5,0.95]$ & $[-0.8,1]$ & $[0,0.2]$ & $1$ & $0.1$ & $[-1.5,1.2]$& $1$ & $1.8$ & $[100,350]$ \\
			\hline
			CC+SN & $71.85\pm 0.36$ & $0.174^{+0.16}_{-0.064} $ & $0.862^{+0.026}_{-0.030}$ & $-0.348 ^{+0.086}_{-0.14}$ & $0.0249^{+0.0051}_{-0.016}$ & $1$ & $0.1$ & $-0.59^{+0.33}_{-0.69}$ & $1$ & $1.8$ & $--$ \\
			\hline
			CC+BAO+DESBAO+SN & $71.76^{+0.32}_{-0.36}$ & $0.15^{+0.16}_{-0.22}$ & $0.851^{+0.027}_{-0.024}$ & $-0.384^{+0.072}_{-0.13}$ & $0.0217^{+0.0051}_{-0.013}$ & $1$ & $0.1$ & $-0.52^{+0.88}_{-0.66}$ & $1$ & $1.8$ & $135.92 \pm 0.95$\\
			\hline
			
			\hline
			
		\end{tabular}
			\caption{{Summary of the $68\%$ confidence limits for each cosmological parameter. The parameters with indicated ranges are those fitted to the data.}}
		\label{tab:data_analysis_table1}
	\end{table}

	\begin{table}[h!]
		\tiny
		\centering
		
		\begin{tabular}{l|c|c|c|c|c|c|c}
			\hline 
			Data & $H_0$ & $\Omega_M$ & $\Omega_{\rm DE}$ & AIC & BIC & $\chi_{\rm red}$ & $r_d$ \\
			\hline
			\multicolumn{8}{c}{\textbf{$\Lambda$CDM}}\\
			\hline 
			\hline
			CC+SN & $73.09 \pm 0.21$ & $0.335 \pm 0.016$ & $0.665$ & $1779.48$ & $1790.39$ & $1.026$ & $--$\\
			\hline
			CC+BAO+DESBAO+SN & $73.32 \pm 0.17$ & $0.315 \pm 0.01$ & $0.685$ & $1811.50$ & $1827.90$ & $1.035$ & $136.38 \pm 0.98$\\
			
			\hline
			\multicolumn{8}{c}{ {\textbf{Model I: $\beta = 2$}}}\\
			\hline
			\hline
			CC+SN & $72.65 \pm 0.30$ & $0.313$ & $0.719$ & $1776.14$ & $1808.88$ & $1.022$ & $--$  \\
			\hline
			CC+BAO+DESBAO+SN & $ 72.61 ^{+0.25}_{-0.32}$ & $0.332$ & $0.708$ & $1805.03$ & $1843.30$ & $1.029$ & $135.63 \pm 0.98$\\
			\hline 
			
			\multicolumn{8}{c}{\textbf{Model II: $\beta =-1$}}\\
			\hline 
			\hline
			CC+SN & $71.85\pm 0.36$ & $0.574$ & $0.773$ & $1768.42$ & $1801.17$ & $1.017$ & $--$\\
			\hline
			CC+BAO+DESBAO+SN & $71.76^{+0.32}_{-0.36}$ & $0.637$ & $0.747$ & $1797.62$ & $1835.89$ & $1.024$ & $135.92 \pm 0.95$\\
			\hline
			\hline
			
		\end{tabular}
		\caption{The statistical summary of Models fitted with cosmological data.}
		\label{tab:Lcdm_model_stat}
	\end{table}
	
	To obtain the Hubble evolution, we solve the autonomous equations, Eq.~(\ref{x_prime})--Eq.~(\ref{s_prime}), by varying them with respect to the time variable \({d}N = d\log(a)\) instead of \({d}\tilde{N}\). The solutions are then used to derive the Hubble evolution, as described in Eq.~\eqref{hubble_evolution_eq}. We varied the initial conditions and parameters related to the field entity while fixing some parameters such as \(\xi_0\), \(s_0\), \(\gamma\), and \(\delta\), with \(\alpha_{1}=1\).
	For the particle production scenario, we explored two values of \(\beta\), corresponding to different models: \(\beta=2\) for Model I and \(\beta=-1\) for Model II.

	\begin{itemize}
		\item \textbf{Model I:} The contour plots for Model I are presented in Fig. [\ref{fig:traingleplot_model1}], and the summary of the best-fit parameters is tabulated in Tab. [\ref{tab:data_analysis_table1}]. The best-fit values from both data sets are nearly identical, demonstrating the model's consistency. Some parameters were held constant as they do not significantly influence the overall evolution of the system. Using these best-fit values, we evaluated the scalar field density and dark matter density, summarized in Tab. [\ref{tab:Lcdm_model_stat}], and also performed a fit with the flat $\Lambda$CDM model for comparison. This allows us to assess the differences in density parameters and how well each model aligns with observational data.
		
		For Model I, the predicted dark matter density is approximately \(\Omega_M \sim 31-33\%\), while the dark energy density is \(\Omega_{\phi} \sim 71\%\). This value is slightly higher than that of the flat-\(\Lambda\)CDM model, where \(\Omega_{\Lambda} \sim 66-68\%\). The Hubble parameter for the current model is around $72.6$ km/s/Mpc, which is slightly lower than the value obtained for the \(\Lambda\)CDM model.
		\begin{figure}[h!]
			\centering
			\subfloat[\label{fig:beta2_pp_cont}]{\includegraphics[scale=0.5]{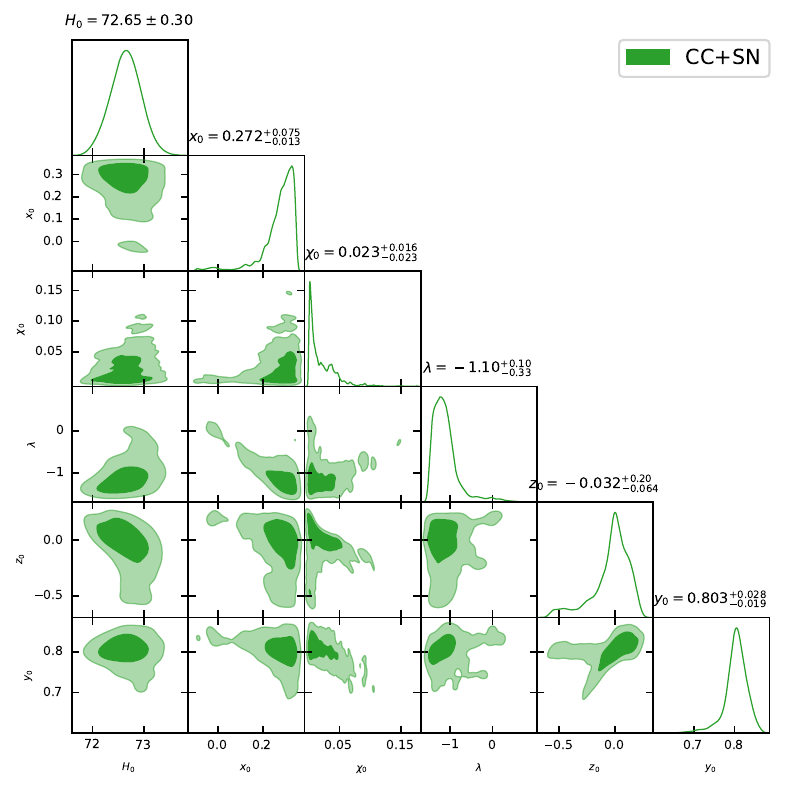}}
			\hspace{0.3cm}
			\subfloat[\label{fig:beta2_all_cont}]{\includegraphics[scale=0.5]{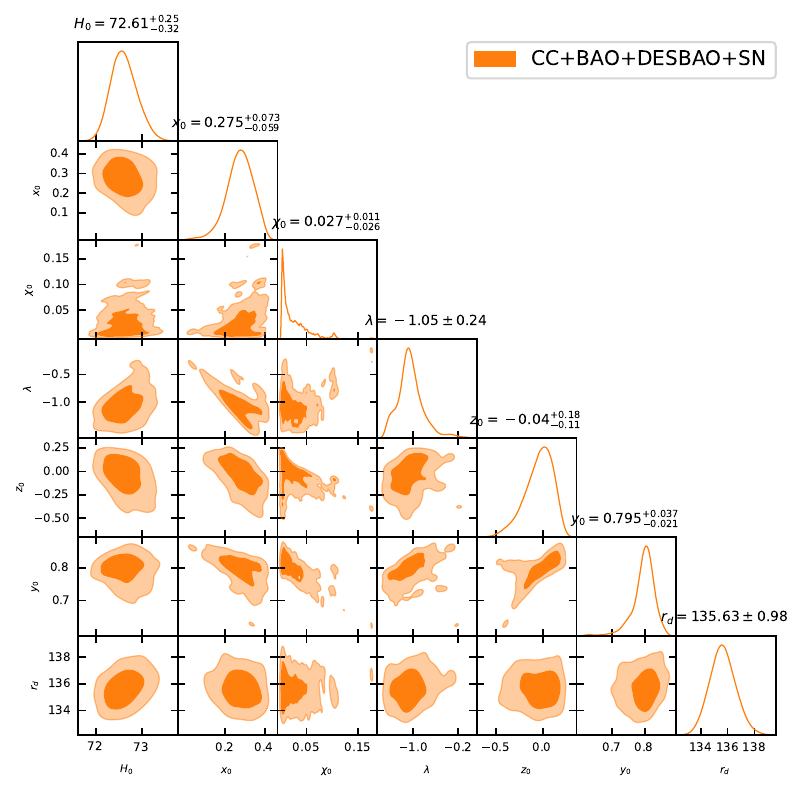}}
			\caption{Contour plot showing the best-fit parameters for Model I.} 
			\label{fig:traingleplot_model1}
		\end{figure}
		
		Additionally, we evaluate determined widely used information criteria to compare the models for evidence, including the Akaike Information Criterion (AIC) and the Bayesian Information Criterion (BIC) \cite{Trotta:2008qt}. The AIC and BIC are defined as:
		\begin{equation}
			\begin{split} 
				\rm{AIC} & = -2 \ln \mathcal{L}_{\rm max} + 2 k,\\
				\rm{BIC}  & = -2 \ln \mathcal{L}_{\rm max} 
				+ k \ln N,
			\end{split}
		\end{equation}
		where \(k\) is the number of independent parameters in the model, \(N\) represents the total number of data points, and \(\mathcal{L}_{\rm max}\) represents the maximum likelihood. To compare each models with \(\Lambda\)CDM, we define the relative difference \(\Delta \text{IC} = \text{IC}(\text{Model}) - \text{IC}(\Lambda \text{CDM})\), where IC refers to the Akaike Information Criterion (AIC) and Bayesian Information Criterion (BIC). The interpretation of \(\Delta \text{IC}\) is as follows: \(\Delta \text{IC} < 2\) indicates substantial support for the model, \(2 < \Delta \text{IC} < 4\) suggests weak support, \(4 < \Delta \text{IC} < 7\) indicates significantly less support, and \(\Delta \text{IC} > 10\) signifies essentially no support for the model \cite{delaCruz-Dombriz:2016bqh,doi:10.1177/0049124104268644,Shi:2023kvu}. Additionally, we report the reduced chi-squared value, \(\chi_{\rm red} \equiv \chi^{2}/N\), where \(N\) represents the degrees of freedom. In general, \(\chi_{\rm red} \approx 1\) indicates a good fit to the data considering the error variance, while \(\chi_{\rm red} > 1\) suggests a poor fit and \(\chi_{\rm red} < 1\) indicates potential overfitting \cite{andrae2010and}.
		
		For Model I, the CC+SN data set yields \(\Delta \text{AIC} = -3.34\) and \(\Delta \text{BIC} = 18.49\). With the CC+BAO+DESBAO+SN dataset, we obtain \(\Delta \text{AIC} = -6.47\) and \(\Delta \text{BIC} = 15.4\). Based on the AIC, Model I, is strongly supported by the data. However, the BIC continues to favor $\Lambda$CDM. 
		
		\item \textbf{Model II:} The contour plots for Model II are shown in Fig. [\ref{fig:traingleplot_model2}], and the best-fit parameter values are summarized in Tab. [\ref{tab:data_analysis_table1}]. This model yields a Hubble parameter of \(H_0 = 71.8\) km/s/Mpc, which is slightly lower than that of Model I. However, a higher dark matter density \(\Omega_M \sim 0.57 - 0.64\) and dark energy density \(\Omega_{\phi} \sim 0.75 - 0.77\) are observed, as shown in Tab. [\ref{tab:Lcdm_model_stat}]. These enhanced fractional energy densities are typical in interacting models, where part of the energy budget contributes to the interaction term which is not directly a physical observable.
		
		{
			It is important to note that the higher value of the matter energy density observed in Model II may also be attributed to the limitations of current late-time data in effectively constraining the parameters of this complex interacting model, particularly due to degeneracies among certain parameters. These degeneracies could potentially be lifted by incorporating early-time observations, such as Cosmic Microwave Background (CMB) data, which is the focus of our forthcoming work.\\
			Previous studies, particularly ref. \cite{Lee:2006za}, have investigated dark sector interactions using a Lagrangian framework and demonstrated that such couplings can significantly influence the positions and amplitudes of the acoustic peaks in the CMB power spectrum. A similar effect may arise in our model, as the interaction alters the evolution of the dark matter fluid. Consequently, a shift in the CMB acoustic peak could be expected. Additionally, as noted in ref. \cite{Lee:2006za}, the coupling can affect the matter power spectrum, with stronger coupling potentially enhancing its amplitude. Such modifications could contribute to alleviating both the $\sigma_8$ and the Hubble tension, as discussed in ref. \cite{Sabogal:2024yha}.
		}
		
		When comparing the statistical benchmarks of this model with \(\Lambda\)CDM, the criteria for the CC+SN data set yield \(\Delta \text{AIC} = -11.06\) and \(\Delta \text{BIC} = 10.78\). For the CC+BAO+DESBAO+SN data set, the differences are \(\Delta \text{AIC} = -13.88\) and \(\Delta \text{BIC} = 7.99\). Based on these results, the AIC criterion strongly supports the current model for both data sets, while the BIC criterion continues to favor \(\Lambda\)CDM. Comparing this model to Model I, the information criterion suggests that Model II offers a notably better fit. 
		
		\begin{figure}[h!]
			\centering
			\subfloat[\label{fig:betam1_pp_cont}]{\includegraphics[scale=0.5]{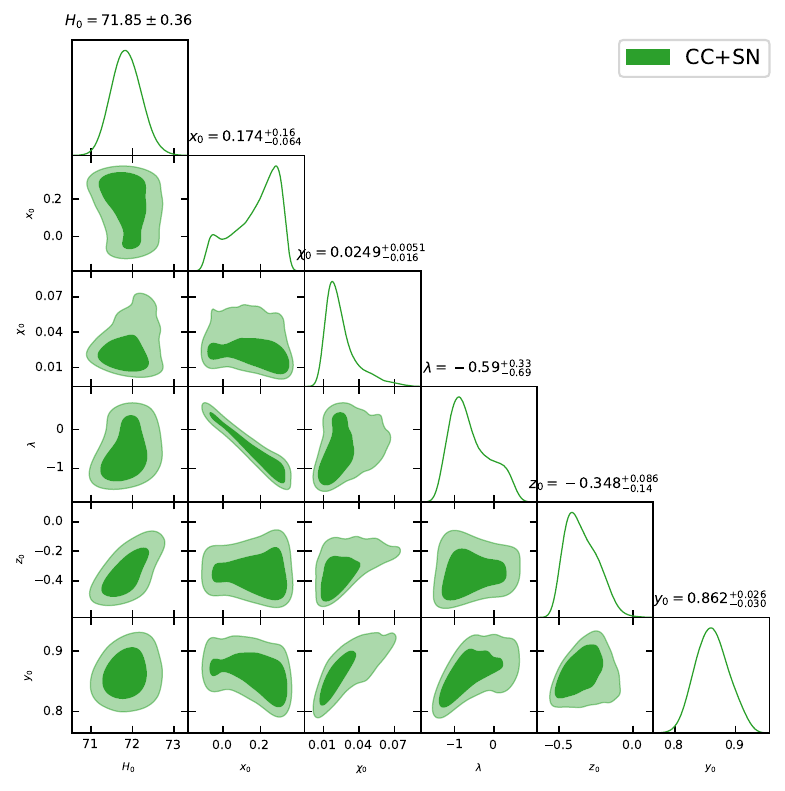}}
			\hspace{0.3cm}
			\subfloat[\label{fig:betam1_all_cont}]{\includegraphics[scale=0.5]{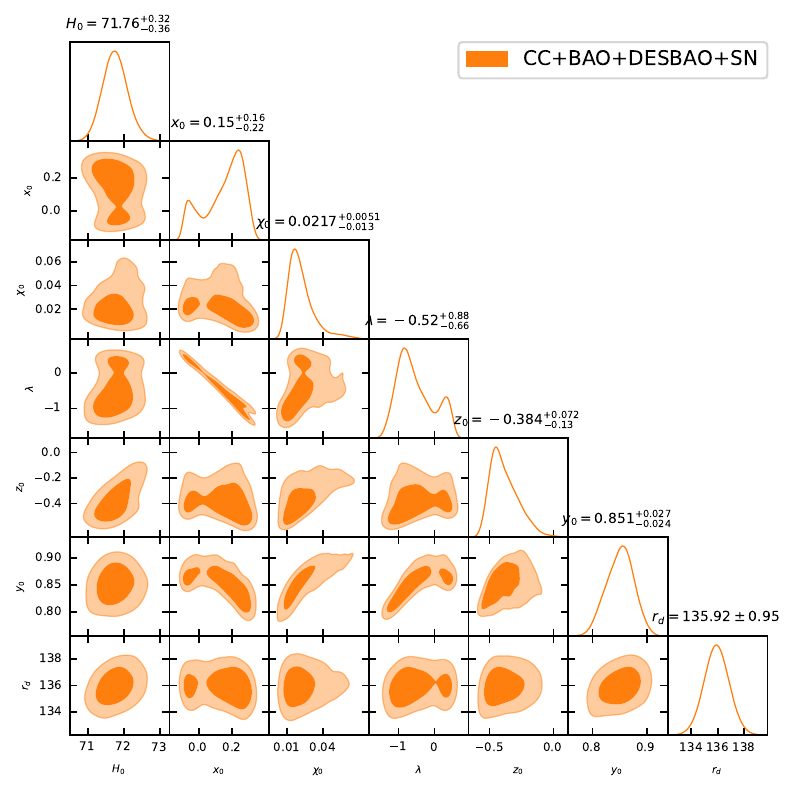}}
			\caption{Marginal distributions of model parameters for Model II. }
			\label{fig:traingleplot_model2}
		\end{figure}
		
		\begin{figure}[t]
			\centering
			\includegraphics[scale=0.7]{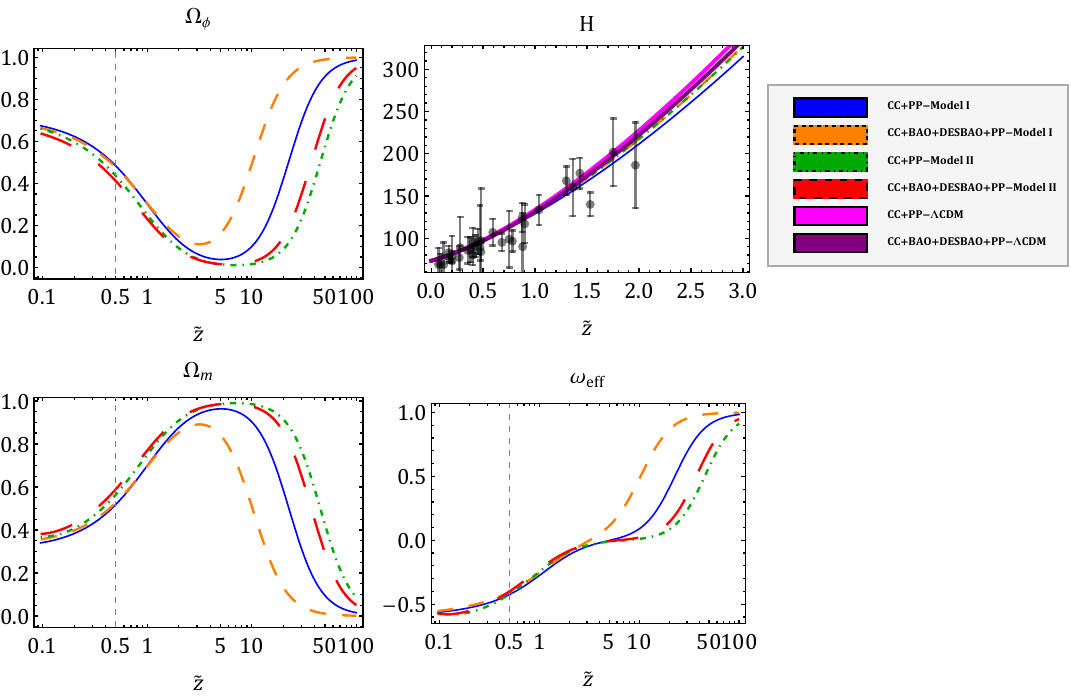}
			
			\caption{Cosmological evolution of physical parameters based on the best-fit values presented in Tab. [\ref{tab:data_analysis_table1}].}
			\label{fig:best_fit}
			
		\end{figure}
		
	\end{itemize}	
	Using the best-fit values obtained for both models, we illustrate the evolution of key cosmological parameters--\(\Omega_M, \ \Omega_\phi, \ \omega_{\rm eff}, \ \& \ H(\tilde{z})\), in Fig. [\ref{fig:best_fit}]. The numerical evolution of the autonomous equations has been calculated using the time variable \(dN = d\log a = -\log(1+\tilde{z})\). In these plots, the colored curves represent different models for distinct data sets. 
	
	In the \(H(\tilde{z})\) plot, we compare the evolution of \(H\) for the current models against \(\Lambda\)CDM. While the evolution is nearly indistinguishable at low redshift \(\tilde{z}<0.5\), deviations become apparent at higher redshifts. The matter fractional energy density dominates (\(\Omega_M \gg \Omega_\phi\)) in the intermediate redshift range \(\tilde{z} \in [0.5, 40]\), during which the effective equation of state (EoS) is approximately zero, indicating a matter-dominated phase. However, at a redshift of \(\tilde{z} \sim 0.5\), the universe transitions into a dark energy-dominated phase as the scalar field fractional energy density starts to exceed that of matter. Consequently, the effective EoS shifts towards negative values, reflecting the onset of dark energy dominance.
	
	When observing the evolution of the energy density parameters, we find that most models display similar behavior at low redshifts. However, distinct features emerge in the blue and orange curves, which represent different data sets. The blue curve indicates a more prolonged matter-dominated phase compared to the orange curve, which corresponds to  Model I. These differences may have non-trivial effects on the matter power spectrum, and further insights could be gained by testing these models with Cosmic Microwave Background (CMB) data.
	
	In contrast, the energy density parameters for Model II show consistent evolution across both data sets, with a relatively extended matter-dominated phase compared to Model I. In the early universe \(\ti{z}>>1\), the scalar field density dominates once again, and the corresponding effective EoS approaches 1, indicating stiff matter behavior—a common feature of quintessence field. During this epoch, the interaction parameter and the matter density parameter diminish, and the scalar field becomes minimally coupled to the dark matter fluid.
}

\section{Conclusion \label{sec:conclusion}}

This paper investigates a non-minimal coupling scenario between a quintessence field and a pressureless fluid using the variational principle. The interaction function \((f)\) is introduced at the Lagrangian level, depending on both field and fluid degrees of freedom. A Lagrange multiplier \(\varphi\), representing the thermodynamic variable related to the gradient of chemical free energy \((F)\) is incorporated into \(f\). This incorporation results in altered equations of motion for the number density and the entropy per particle. The interaction function \((f)\) serves as the source term for the production or annihilation of dark matter particles and entropy generation, consequently altering the evolution of the fluid's temperature. Assuming the dark matter fluid behaves as a pressureless ideal gas, the entropy is found to be a logarithmic function of temperature. Because of non-minimal coupling via \(f\), the covariant derivatives of the individual stress-energy tensors are no longer conserved, facilitating the flow of energy from the field to the fluid through the field derivative of the interaction function \(f_{,\phi}\dot{\phi}\).

The equations of motion for both the field and fluid are altered by the non-minimal coupling. The dynamics of the composite system require assuming a particular form of the interaction function, which depends on several field-fluid parameters, resulting in multiple free parameters in the model. Constraining these model parameters is essential and has been explicitly discussed using the dynamical system stability technique. For this study, we adopt an exponential interaction function and an exponential potential for the quintessence field.

The stability of the system is analyzed using the linearization technique in the six-dimensional phase space, yielding 13 critical points categorized based on their physical characteristics to describe different phases (non-accelerating to accelerating) of the universe. Two models were particularly studied where the interaction function is \(f \propto \rho^{2}\) and \(f \propto \rho^{-1}\). For these models, the parameters were constrained by evaluating the physical viability of the critical points. The physically viable critical points are those that generate accelerating or non-accelerating solutions, with field and fluid densities bound between \(0\) to \(1\). Additionally, these points must yield positive values for thermodynamic variables, such as the number density \((\chi)\), the temperature variable \((\xi)\), and entropy per particle density \((s)\). It was found that for some critical points, the linearization technique is insufficient for stability analysis due to vanishing eigenvalue. Given the phase space dimension is more than 3, using mathematical tools like the center manifold theorem becomes extremely challenging. Therefore, the stability of such points was investigated by numerically evolving the system near their existence condition using randomized initial conditions.

	To assess the viability of the proposed models, we performed a detailed analysis using two different data combinations: (i) cosmic chronometers (CC) + supernovae (SN), and (ii) CC + SN + Baryon Acoustic Oscillations (BAO) + Dark Energy Spectroscopic Instrument (DESBAO) data. Here, we treated \(r_d\) as a free parameter, and applied joint chi-squared analyses for both data combinations. Two different models were analyzed: Model I with \(\beta = 2\), and Model II with \(\beta = -1\), which describe different interaction terms between dark matter and dark energy.

	For the first combination (CC+SN), the best-fit parameters for Model I resulted in \(H_0 = 72.65\) km/s/Mpc, \(\Omega_M = 31\%\), and \(\Omega_{\phi} \sim 72\%\), showing a modest deviation from \(\Lambda\)CDM. With the second combination (CC+SN+BAO+DESBAO), Model I predicted a slightly higher matter density parameter compared to \(\Lambda\)CDM. Model II, in contrast, exhibited a higher dark matter density of \(\Omega_M \sim 57\%-64\%\) and a slightly lower \(H_0 = 71.8\) km/s/Mpc than both Model I and \(\Lambda\)CDM. Despite these differences, both models presented a similar qualitative evolution of cosmological parameters, including fractional energy densities \(\Omega_M\), \(\Omega_\phi\), and the effective equation of state \(\omega_{\rm eff}\). While both models mimic \(\Lambda\)CDM at low redshifts, they deviate more significantly at higher redshifts. Specifically, Model I shows a shorter matter-dominated phase, potentially impacting cosmic structure growth and the matter power spectrum. Model II, with a more prolonged matter-dominated phase, outperformed Model I statistically.
	
	To evaluate these models, we applied the Akaike Information Criterion (AIC) and the Bayesian Information Criterion (BIC). AIC analysis demonstrated strong support for both models compared to \(\Lambda\)CDM, particularly with the combined data sets. However, BIC values suggested that while Model I showed some support, it was less compelling than \(\Lambda\)CDM. Model II, by contrast, exhibited better fits according to both AIC and BIC, especially in the CC+BAO+DESBAO+SN combination, although \(\Lambda\)CDM remained competitive under the BIC criterion.
	
	In conclusion, although both models are consistent with current observational data, the interaction between dark matter and dark energy in Model II provides a more compelling alternative—particularly when combined datasets are considered.
	Future work will explore a non-minimal coupling between the scalar field and baryonic matter or radiation in a perturbed cosmological background. We also plan to include CMB data to impose tighter constraints on the model and evaluate its ability to alleviate the Hubble tension.

\section*{Acknowledgements}
The author acknowledges the support of National Natural Science Foundation of China under Grants No. W2433018 and No. 11675143, and the National Key Reserach and Development Program of China under Grant No. 2020YFC2201503.

  \bibliographystyle{elsarticle-num} 
  \bibliography{reference}

\end{document}